\documentclass [10pt, twocolumn,superscriptaddress,showpacs] {revtex4}
\usepackage {amsmath}
\usepackage {graphicx}
\newcommand{\beq}{\begin{equation}}
\newcommand{\eeq}{\end{equation}}
\newcommand{\bea}{\begin{eqnarray}}
\newcommand{\eea}{\end{eqnarray}}
\newcommand{\be}{\begin{equation}}
\newcommand{\ee}{\end{equation}}

\begin {document}

\title {Quantum-critical pairing in electron-doped cuprates}
\author{Yuxuan Wang and Andrey Chubukov}
\affiliation{Department of Physics, University of Wisconsin, Madison, WI 53706, USA}
\date{\today}
\begin{abstract}
We revisit the problem of spin-mediated superconducting pairing  at the antiferromagnetic quantum-critical point  with the ordering momentum $(\pi,\pi)=2{\bf k}_F$.
The problem has been previously considered by  one of the authors [P. Krotkov and A. V. Chubukov, Phys. Rev. Lett. {\bf 96},  107002 (2006); Phys. Rev. B {\bf 74}, 014509 (2006)]. However, it  was later pointed out [D. Bergeron {\it et al.}, Phys. Rev. B {\bf 86}, 155123 (2012)] that  that analysis neglected umklapp processes for the spin polarization operator. We incorporate umklapp terms and re-evaluate the normal state self-energy and the critical temperature of the pairing instability. We show that the self-energy has a Fermi-liquid form and obtain the renormalization of the quasiparticle residue $Z$, the Fermi velocity, and the curvature of the Fermi surface. We argue that the pairing is a BCS-type problem, but go one step beyond the BCS theory and
 calculate the critical temperature $T_c$ with the prefactor.  We apply the results to electron-doped cuprates near optimal doping and obtain  $T_c \geq  10 \rm K$, which matches the experimental
  results quite well.
\end{abstract}
\pacs{74.20.Mn, 74.20.Rp}
\maketitle

\section{Introduction}

Superconductivity near a quantum-critical point (QCP) in a metal is one of the most studied and debated topics in the physics of strongly correlated electron
systems~\cite{subir_book,piers_andy,acs,tremblay}. When a metal is brought, by doping, external field, or pressure, to a near vicinity of a phase transition into a state with either
spin or charge density-wave order,
 corresponding collective bosonic excitations become soft and fermion-fermion interaction, mediated by collective bosons, gets enhanced. Quite often, this
 interaction turns out to be attractive in one or the other pairing channel, and
  gives rise to enhanced superconducting $T_c$ near a QCP.
   Examples of the pairing mediated by soft bosonic fluctuations include the pairing
   in double-layer composite fermion metals~\cite{nick}, pairing due to singular momentum-dependent interaction~\cite{extra}, color
    superconductivity of quarks mediated by gluon exchange~\cite{color}, pairing by singular gauge fluctuations~\cite{gauge,aim_el,ms_1},
  pairing by near-critical ferromagnetic and antiferromagnetic spin fluctuations~\cite{fm_qcp,afm_qcp,acf,lonz-pines,ms_2,efetov,am_qcp,moon_sa},
   and phonon-mediated pairing at vanishing Debye frequency~\cite{phonons}.

The pairing at a QCP in dimensions $D \leq 3$ is generally a non-trivial problem because near a QCP fermions do not display a conventional Fermi-liquid (FL)
behavior down to the smallest frequencies, at least in some (hot) regions on the Fermi Surface (FS).  To obtain the pairing instability in this situation one has to go
beyond the leading logarithmical approximation as the summation of the leading logarithms does not lead to an instability~\cite{acf,wang}, except for a special case of a color superconductivity~\cite{color}.

   Pairing near the ${\bf Q} = (\pi,\pi)$ antiferromagnetic QCP attracted most of the attention in recent  years, particularly in $D=2$,  because of its
  relation to $d$-wave superconductivity in the cuprates.~\cite{acs,lonz-pines,ms_2,efetov,scal_1,scal,moria,manske,ramisak,FLEX,X,mar_fit}.  The FS in hole-doped cuprates 
 around optimal doping
is an open electron FS (closed hole FS) which contains four pairs of hot spots (points for which ${\bf k}_F$ and ${\bf k}_F + {\bf
  Q}$ are both on the FS).  The hot spots are located near $(0,\pi)$ and other symmetry-related points.  At each hot spot, fermionic self-energy at a QCP has
  a non-FL form $\Sigma (\omega) \propto \omega^{a}$ with $a \approx 1/2$, down to the lowest frequencies~\cite{millis_1/2,acs,ms_2,senthil}.
  The pairing of these hot fermions, the relative role of quasiparticles with non-FL and FL forms of the self-energy,  and the interplay between the pairing
  and the bond-order instabilities near a QCP are intriguing phenomena which are not fully understood yet, but for which a strong progress has been made theoretically in
  the last few years~\cite{ms_2,efetov,wang}.

   The present work is devoted to the analysis of superconductivity at the antiferromagnetic QCP, but in the special case when ${\bf Q}$ coincides with
   $2{\bf k}_F$ along the diagonals in the Brillouin zone (see FIG. \ref{fs})~\cite{aim_el}.
    In this situation, there are only two pairs of hot spots, located along the two diagonals of the
   Brillouin zone, and for each pair Fermi velocities at ${\bf k}_F$ and ${\bf k}_F + {\bf Q}$ are strictly antiparallel.
 This case is applicable to
    electron-doped cuprates ${\rm Nd}_{2-x}{\rm Ce}_x{\rm Cu}_3{\rm O}_7$ and  ${\rm Pr}_{2-x}{\rm Ce}_x{\rm Cu}_3{\rm O}_7$~\cite{el_rmp}.
  The FS of these 
    materials near the onset of spin density wave (SDW) order has four pairs of 
    hot spots located close enough to zone  diagonals. This makes our model a  good starting point. As $T_c$ only increases as hot spots move apart (see below), our result places the lower boundary on $T_c$ in these materials.
    
    The phase diagram of electron doped cuprates is somewhat better understood than that of hole-doped materials in the sense that the pseudogap physics of these materials is most likely due to magnetic precursors~
    \cite{tremblay_1,el_rmp,zimmers,koitzsch03,Aiff03,greven}.
     This in turn implies that magnetic fluctuations are strong and well may be relevant to superconductivity.
    An earlier RPA study has found~\cite{onufr} that the electron doping, at which magnetic order emerges, is close to the one at
    which hot spots merge along zone diagonals. At larger electron doping, there are no hot spots and no magnetism. At smaller dopings,  magnetic order
    emerges and, simultaneously, each pair of diagonal hot spots splits into two pairs which move away from diagonals in two different directions.

The quantum-critical pairing near a $2{\bf k}_F$ QCP has been earlier considered by one of us in Refs.~\cite{krotkov,dan}. It was found
that there is a sizable attraction in the $d_{x^2-y^2}$ channel, despite that at $2{\bf k}_F$ QCP the strongest interaction connects the points where
$d_{x^2-y^2}$ gap vanishes. This result stands. However, the value of $T_c$   has to be reconsidered because the analysis in~\cite{krotkov,dan}
  used for the normal state self-energy at the hot spots the non-FL form  $\Sigma (\omega) = \omega^{3/4}$, which was obtained in \cite{krotkov} and in
  earlier work~\cite{aim_el} by neglecting umklapp processes.  Recent study~\cite{subir_el} has found that umklapp processes  severely reduce the
  self-energy, such that quasiparticles remain coherent even at a QCP. The imaginary part of the self-energy at the QCP
  scales as $\omega^{3/2} \log \omega$, i.e., the Fermi liquid is non-canonical in the notations of ~\cite{maslov}, but, still,
  ${\rm Im} \Sigma \ll \omega$ at small frequency, and the Fermi liquid criteria of  long-lived quasiparticles near the FS remains valid.
  Whether the $\omega^{3/2}$ self-energy is responsible for the observed $T^{1.6}$ behavior of resistivity in overdoped
  La$_{2-x}$Ce$_{x}$CuO$_{4}$ (Ref. [\onlinecite{rick}]) remains to be seen.

 In this paper, we reconsider the problem of the pairing at the $2{\bf k}_F$ QCP using the correct form of the self-energy.  We show that the
  pairing is a FL phenomenon, i.e., it is fully determined by the coherent component of the quasiparticle Green's function
    and depends on the quasiparticle $Z$ factor and the effective mass.
    Moreover, when spin-fermion coupling is small compared to
   the Fermi energy, the pairing can be analyzed within the weak coupling, BCS-type analysis. Our goal is to go one step beyond
   the BCS theory and compute $T_c$ exactly, with the numerical prefactor.

   The calculation we present here  is similar in spirit to the one done in early days of BCS superconductivity for a model of fermions with a constant attractive
    interaction~\cite{gorkov-melik}, but is more involved, as in our case the pairing interaction contains the propagator of low-energy collective bosons which
    strongly depends on the transferred momentum and the transferred frequency.  We show that these two dependencies make calculations
    somewhat tricky, but still doable.

 We consider the low-energy model of fermions located near Brillouin zone diagonals and assume that fermions interact by exchanging  near-critical soft collective fluctuations in the spin
 channel (the spin-fermion model). The model contains two parameters: the overall energy scale ${\bar g}$, which is the effective
fermion-fermion interaction mediated by 
 spin fluctuations,
   and the dimensionless coupling $\lambda$, which determines mass renormalization and the renormalization of the quasiparticle $Z$ factor~\cite{acs,ms_2}. In our case,
    the coupling $\lambda$ is one-third power of the  ratio of ${\bar g}$ and the effective Fermi energy of
    quasiparticles near Brillouin zone diagonal:
      \be
    \lambda = \left(\frac{{\bar g}}{4\pi E_F}\right)^{1/3},
    \label{ch_2}
    \ee
   We approximate the fermionic dispersion near Brillouin zone diagonals by
   $\epsilon_k = v_F (k_x +  \kappa k^2_y/2)$, where $k_x$ and $k_y$ are the directions along and transverse to zone diagonals, measured relative to a hot
   spot, $v_F$ is the Fermi velocity along zone diagonal and $\kappa$ is the curvature of the FS. In terms of these parameters
    $E_F = v_F/(2\kappa)$.

 In the normal state we found that the quasiparticle residue $Z$, the  Fermi velocity $v_F$, and the curvature $\kappa$  differ from free-fermion values by
 corrections of order  $\lambda$:
  \be
 Z = 1- 0.7 \lambda, ~~v^*_F = v_F (1+0.05 \lambda),~~\kappa^* = \kappa(1-1.45 \lambda),
 \label{fr_2_2}
 \ee
 Observe that the renormalizations of $Z$ and $v^*_F$ do not satisfy $Z v_F/v^*_F=1$, which holds in   Eliashberg-type theories (ET's) of
  electron-phonon~\cite{phonons} and electron-electon~\cite{acs,acf} interactions.
A similar result was
obtained in the calculation of self-energy in the AFM state
\cite{das2}.
ramisek
The reason for the discrepancy with ET's is quite fundamental: in ET's
    the self-energy $\Sigma ({\bf k}, \omega_m) \approx i\lambda \omega_m$  depends only on frequency and comes from intermediate fermions with
     energies comparable to $\omega_m$, other corrections are relatively small in Eliashberg parameter ($(\omega_D/E_F)^{1/2}$ for electron-phonon
     interaction).  In our case, Eliashberg parameter is of order one, and
     there are two contributions to $\Sigma ({\bf k}, \omega_m)$ of comparable strength. One comes from intermediate fermions with energies
      comparable to $\omega_m$ and scales as $\omega_m$, i.e., depends only on frequency. The other comes from intermediate fermions with energies much larger than $\omega_m$,
       and scales as $i\omega_m + v_F(k-k_F)$.  Because the two contributions are of the same order, $\partial \Sigma/\partial (i\omega_m)$ and $(1/v_F)
       \partial \Sigma/\partial k$ are comparable. As a result, the renormalizations of $Z$ and of Fermi velocity are comparable in strength (both are of
       order $\lambda$), but the prefactors  are different.

  We used normal state results as an input for the pairing problem
   and computed superconducting $T_c$ at the QCP.
   We found
   \be
   T_c = 0.0013
  ~ \frac{\bar g} {\lambda}~ e^{-\frac{C}{\lambda}}, ~~C = 0.6874
   \label{ch_1}
   \ee
   The exponential dependence on $\lambda$ and the overall factor
    ${\bar g}/\lambda$  can be obtained within logarithmical BCS-type treatment. However,
    to obtain the overall  numerical factor in (\ref{ch_1}), we had to go beyond the logarithmical approximation and use the fact that theramisek
    boson-mediated interaction is dynamical and decays at large frequencies.  This part of our  analysis is similar to the
    calculation of $T_c$ in ETs~\cite{phonons,acf}. However, as we said, the similarity is only partial because
  the renormalized Green's function in our case is different from the one in ET.
      In another distinction from ET, in our case  we need to include into consideration non-ladder diagrams which account for Kohn-Luttinger-type
      renormalization of the pairing interaction~\cite{kl}. These diagrams do not affect the exponent but contribute $O(1)$
      to the numerical prefactor in $T_c$.  In ET theories, the contribution from Kohn-Luttinger renormalizations is small in Eliashberg parameter.

\begin{figure}[htbp]
\includegraphics[width=\columnwidth]{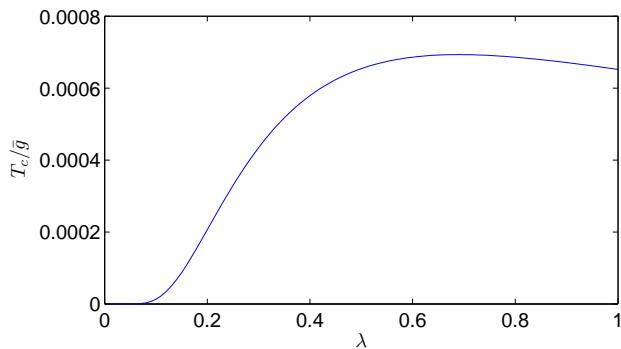}
\caption{The critical temperature $T_c/\bar{g}$ as a function of the coupling $\lambda$.}
\label{fig8}
\end{figure}

We plot $T_c/{\bar g}$  as a function of $\lambda$ in  FIG. \ref{fig8}.
We see that $T_c$ increases with $\lambda$ at small $\lambda$,  passes through a broad maximum
   at $\lambda \sim 0.5$ and slowly decreases at larger $\lambda$.
   The actual value of $\lambda$ can be extracted  from the data.
 The energy ${\bar g}$ can be deduced from optical measurements
   in the magnetically ordered Mott-Heisenberg state at half-filling~\cite{optics_el} where ${\bar g}$ coincides with the optical gap.
    The data yield  ${\bar g} \sim 1.6-1.7 \rm eV$, which is essentially the same as in
   hole-doped cuprates (this scale is close to the charge-transfer energy $U$ in the effective Hubbard model).
The Fermi velocity $v_F$ and the curvature $\kappa$   can be extracted from the
  ARPES measurements of optimally doped ${\rm Nd}_{2-x}{\rm Ce}_x{\rm Cu}_3{\rm O}_7$~\cite{exp_el}. The fit yields
  $v_F = 0.87 \rm eV$, $\kappa = 0.31$, and
  the effective Fermi energy  $E_F  = v_F/(2\kappa)\sim 1.4 \rm eV$. Substituting into (\ref{fr_2_2})  we obtain
  $\lambda \sim 0.46$. For such $\lambda$, $Z \approx 0.7$, i.e.,  weak coupling approximation is reasonably well justified.
 Using $\lambda =0.46$ and ${\bar g} =1.7 \rm eV$,  we obtained
    $T_c = 0.0006 {\bar g} \sim 12  \rm K$. This value is actually the lower theoretical boundary on $T_c$ by two reasons. One is theoretical --
    we found that $T_c$  get enhanced if we keep the fermionic bandwidth (the upper limit of the low-energy theory) finite. The second is practical --  in optimally doped
     ${\rm Nd}_{2-x}{\rm Ce}_x{\rm Cu}_3{\rm O}_4$ and  ${\rm Pr}_{2-x}{\rm Ce}_x{\rm Cu}_3{\rm O}_4$  hot spots are  located close to, but still at some distance from zone diagonals~\cite{el_rmp}. When they are far apart, $T_c$ is much higher~\cite{wang,norman_1}, and it is natural to expect that $T_c$
      gets higher when hot spots split.ramisek
       Another feature of real materials, which may affect $T_c$ a bit, is the observation~\cite{greven} that SDW antiferromagnetism and superconductivity
        in ${\rm Nd}_{2-x}{\rm Ce}_x{\rm Cu}_3{\rm O}_{4+\delta}$ may be actually separated by weak first-order transition rather than co-exist~\cite{dagan04}, in which case the magnetic
         correlation length remains large but finite at optimal doping.  Still, the value which we found theoretically is in quite reasonable agreement with the experimental $T_c \sim 10-25 \rm K$ in ${\rm Nd}_{2-x}{\rm Ce}_x{\rm Cu}_3{\rm O}_4$ and  in ${\rm Pr}_{2-x}{\rm Ce}_x{\rm Cu}_3{\rm O}_4$ (Refs.~\cite{takagi,TC2}).
  Also, our $\lambda =0.46$ is close to the position of the broad maximum of $T_c (\lambda)$  in FIG. \ref{fig8}, hence
  $T_c$ only slightly increases or decreases if
    the actual $\lambda$ differs somewhat from our estimate.

The paper is organized as follows. In the next section we introduce the spin-fermion model as the minimal model to describe interacting fermions near the
$2{\bf k}_F$ spin-density-wave instability. In Sec. III we present
 one-loop normal state calculations, which include umklapp scattering.  We obtain
 the spin polarization operator, which accounts for the dynamics of collective excitations, and use it to obtain fermionic self-energy to first order in $\lambda$.   In Sec. IV we solve for $T_c$
  in one-loop (ladder)
 approximation. We show that $T_c$ is exponentially small in $\lambda$ and find the prefactor to one-loop order.
In Sec. V we analyze two-loop corrections to $T_c$ and show that they  change the numerical prefactor for $T_c$ by a finite amount. We argue that  higher-loop corrections are irrelevant as they only change $T_c$ by a factor $1 + O(\lambda)$. Combining one-loop and two-loop results, we obtain the  full expression
 for $T_c$ at weak coupling,  Eq. (\ref{ttcritical_2}).
  We also discuss in Sec. V the effect on $T_c$ from  lowering the upper energy
 cutoff $\Lambda$ of the theory. When $\Lambda$ is of order bandwidth, $W$, $T_c$ is essentially independent on $\Lambda$ as long as $T_c \ll W$.
  However, if, by some reasons,
  $\Lambda$ is smaller that $W$, $T_c$ gets larger, and its increase becomes substantial when $\Lambda < {\bar g}^{1/3} W^{2/3}$.
  We compare our results with the experiments in Sec. VI and summarize our results in Sec. VII.

\section{The model}

\begin{figure}[htbp]
\includegraphics[width=.55\columnwidth]{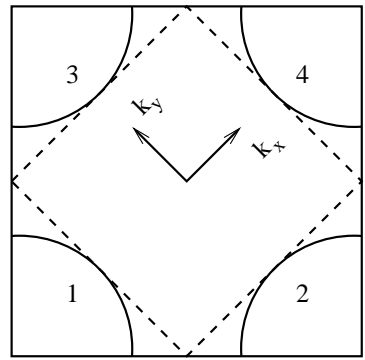}
\caption{Fermi surface at a particular electron doping when the Fermi surface touches the magnetic Brillouin zone  along the zone diagonals.
 The four diagonal Fermi surface points (hot spots) are labeled by numbers.
We assume that this doping is close to the one at which  antiferromagnetic instability emerges at momentum ${\bf Q}=(\pi, \pi)$.}
\label{fs}
\end{figure}

We consider fermions on a square lattice at a density larger than one electron per cite (electron doping) and use tight-binding form of
electron dispersion with hopping to first and second neighbors.  We choose doping at which free-fermion FS
touches the magnetic Brillouin zone along the diagonals, as shown in FIG. \ref{fs}.  We assume, following earlier works~\cite{onufr}, that at around this doping the system
 develops a spin-density-wave (SDW) order with momentum ${\bf Q}=(\pi, \pi)$.
 For convenience, throughout the paper we will use a rotated reference frame with  $(k_x, k_y)$ shown in FIG. \ref{fs}. In this frame, the SDW
 ordering momentum is $(0, \sqrt{2}\pi)$ or $(\sqrt{2}\pi,0)$.

  We analyze the physics near the ${\bf Q}=2{\bf k}_F$ antiferromagnetic QCP within
  the semi-phenomenological spin-fermion model~\cite{acs,acf,ms_2,efetov,aim_el,krotkov,subir_el}. The model assumes that antiferromagnetic correlations
  develop already at
   high energies, comparable to bandwidth, and mediate interactions between low-energy fermions.
   In the context of superconductivity, spin-mediated interaction then plays the same role of a pairing glue as phonons do in conventional superconductors.
  The static part of the spin-fluctuation propagator comes from high-energies and should be treated as an input for low-energy theory. However, the
   dynamical part of the propagator should be self-consistently obtained  within the model as it
     comes entirely from low-energy fermions.

     The
    Lagrangian of the model is
    given by~\cite{acs,ms_2,efetov}
 \begin{align}
{\mathcal{S}} =&-\int_{k}^{\Lambda }G_{0}^{-1}\left( k\right) \psi _{k,\alpha }^{\dagger }\psi _{k,\alpha }+\frac{1}{2}\int_{q}^{\Lambda
}\chi _{0}^{-1}\left( q\right) \ {\bf{S}}_{q}\cdot {\bf{S}}_{-q}  \nonumber \\ &+g\int_{k,q}^{\Lambda }\psi _{k+q,\alpha }^{\dagger }\sigma
_{\alpha \beta }\psi _{k,\beta }\cdot {\bf{S}}_{-q}.\
  \label{startac}
\end{align}
  where $\int_k^\Lambda$ stands for the integral over $d-$dimensional
 ${\bf{k}}$ (up to some upper cutoff $\Lambda$) and the sum runs over fermionic and bosonic
  Matsubara frequencies. The $G_{0}\left( k\right) = G_0 ({\bf k},\omega_m) = 1/(i\omega_m - \epsilon_{\bf k})$
  is the bare  fermion propagator and
   $\chi _{0}\left( {\bf q}+{\bf Q}\right) = \chi_0/({\bf q}^2 + \xi^{-2})$
  is the static propagator of collective bosons, in which $\xi^{-1}$ measures the distance to the QCP, and ${\bf q}$ is measured with respect to ${\bf Q}$.
  At the QCP,  $\xi^{-1} =0$.
   The fermion-boson coupling $g$ and $\chi_0$ appear in theory only in combination ${\bar g} = g^2 \chi_0$, which has the dimension of energy.

The interaction between fermions and collective spin bosons gives rise to fermionic self-energy $\Sigma ({\bf k}, \omega_m)$, which modifies
 fermionic propagator to $G^{-1} ( {\bf k}, \omega_m) = G^{-1}_0 ({\bf k}, \omega_m) + \Sigma ({\bf k}, \omega_m)$, and bosonic self-energy which modifies bosonic propagator to
 $\chi^{-1}\left({\bf q}+{\bf Q}, \Omega_m)\right) = \left({\bf q}^2 + \xi^{-2} + \Pi \left({\bf q}+{\bf Q}, \Omega_m\right)\right)/\chi_0$.
  We focus on the hot regions
  on the FS, which are most relevant to the pairing when ${\bar g}$ is smaller than the cutoff $\Lambda$. In our case, there are four hot regions around
   Brillouin zone diagonals, which
   we labeled $1$ to $4$ in FIG. \ref{fs}.
    The physics in one hot region
    is determined by the interaction with another hot region,  separated by ${\bf Q}$. This creates two ``pairs" -- (1, 4), and (2, 3).
     However, the pairs cannot be fully separated because
     ${\bf Q}$ and
    $-{\bf Q}$ differ by inverse lattice vector, hence  umklapp processes arramiseke allowed~\cite{subir_el}.  As a result, the interaction between fermions in
    regions
     $1$ and $4$ is mediated by $\chi ({\bf q}+{\bf Q}, \Omega_m)$ whose polarization operator $\Pi ({\bf q}+{\bf Q}, \Omega_m)$ has contribution from   
     fermions  in the same regions 1 and 4, but also from
      fermions in the regions 2 and 3.

    We define fermion momenta relative to their corresponding hot spots. The dispersion of a fermion is linear in transverse momentum (the one along the
    Brillouin zone diagonal)
     and quadratic in the momentum transverse to the diagonal. 
     Specifically, the dispersion relation in region 1 is
   \begin{align}
\epsilon_k=-v_F\left(k_x+ \kappa \frac{k_y^2}{2}\right),
\end{align}
where $\kappa$ is
 the curvature of the FS.

\section{Normal State Analysis}

 The spin polarization operator $\Pi ({\bf q}+{\bf Q}, \Omega_m)$  is the dynamical part of the particle-hole bubble with external momentum near ${\bf Q}$. The dressed
 dynamical spin fluctuations  give rise to fermionic self-energy $\Sigma ({\bf k}, \omega_m)$, which in turn affects the form of $\Pi ({\bf q}+{\bf Q}, \Omega_m)$. This mutual
 dependence
  generally
    implies that $\Sigma ({\bf k}, \omega_m)$ and $\Pi({\bf q}+{\bf Q}, \Omega_m)$ form a set of two coupled equations.
   When hot spots are far from zone diagonals and Fermi velocities at ${\bf k}_F$ and ${\bf k}_F + {\bf Q}$ are not antiparallel (like in
   hole-doped cuprates),
    the two equations decouple because the bosonic polarization operator has the  Landau damping form $\Pi ({\bf q}+{\bf Q}, \Omega_m) = \gamma |\Omega_m|$, and the
     prefactor $\gamma$ does not depend on fermionic self-energy
      as long as the latter predominantly depends on frequency.
       Evaluating fermionic $\Sigma$ with the
        Landau overdamped $\chi ({\bf q}+{\bf Q}, \Omega_m)$ one can in turn verify~\cite{acs,ms_2} that $\Sigma$  predominantly depends on frequency near a QCP, i.e.,
         equations for $\Sigma ({\bf k}, \omega_m) \approx \Sigma (\omega_m)$ and $\Pi ({\bf q}+{\bf Q}, \Omega_m)$ do indeed decouple.
       This decoupling allows one to compute the Landau damping using free-fermion propagator, even when $\Sigma (\omega_m)$ is not small,
       and  use the dynamical $\chi ({\bf q}+{\bf Q}, \Omega)$ with Landau damping term
       in the  calculations of the fermionic self-energy~\cite{comm_a}
     In our case, the dynamical part of $\chi$ is not  Landau damping because Fermi velocities at ${\bf k}_F$  and ${\bf k}_F + {\bf Q}$
      are strictly antiparallel, and $\Pi ({\bf q}+{\bf Q}, \Omega_m)$ does depend on fermionic self-energy~\cite{dan}.  This generally requires full
      self-consistent analysis of the coupled set of non-linear equations for $\Sigma ({\bf k}, \omega_m)$ and $\Pi ({\bf q}+{\bf Q}, \Omega_m)$.
      Fortunately, in our case the system preserves a FL behavior even at the QCP, and for ${\bar g} < W$, which we assume to hold,
       calculations can be done perturbatively rather than self-consistently.
         Below we will obtain lowest-order (one-loop) expressions for the polarization bubble and fermionic self-energy and
         use them to compute $T_c$ in the ladder approximation.  We then discuss
       how these expressions are affected by two-loop diagrams.

\subsection{One-loop bosonic and fermionic self-energies}

\begin{figure}[htbp]
\includegraphics[width=.8\columnwidth]{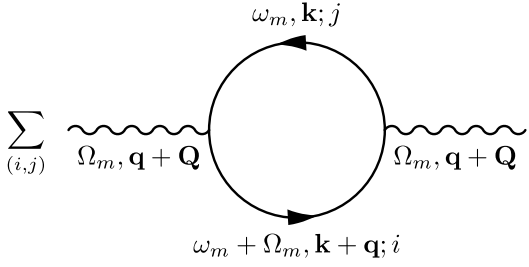}
\caption{The one-loop polarization bubble. Labels $i$ and $j$ denotes fermions at different hot spots. The full expression is the sum of
  direct and umklapp processes, $(1,4), (4,1)$, $(2,3), (3,2)$.}
\label{RPA}
\end{figure}

 The one-loop Feynman diagram for the polarization operator is shown in FIG. \ref{RPA}.
The  polarization operator contains contributions from {\it direct} and {\it umklapp} scattering between hot spots separated by either ${\bf Q}$ or $-{\bf Q}$ (the combinations of $(i,j)$ of internal fermions can be $(1,4), (4,1), (2,3),$ and $(3,2)$. For external momenta near
${\bf Q} = (\pi,\pi)$,
  the
 processes $(1,4)$ and $(4,1)$ are  direct and $(2,3)$ and $(3,2)$ are umklapp.
  Only direct processes have been considered in Ref. [\onlinecite{aim_el,krotkov}], but, as was pointed out in \cite{subir_el}, all four processes should be  included into $\Pi$. The  authors of Ref. [\onlinecite{subir_el}] obtained
 \begin{align}
  \Pi({\bf q} + {\bf Q} ,\Omega)=& \Pi^{(0)}(q_x,q_y,\Omega)  +\Pi^{(0)}(-q_x,q_y,\Omega) \nonumber \\
+&\Pi^{(0)}(q_y,q_x,\Omega)+\Pi^{(0)}(-q_y,q_x,\Omega),
  \end{align}
   where~\cite{aim_el,krotkov}

\begin{align}
\Pi^{(0)}(q_x,q_y,\Omega)=\frac{{\bar g}}
{\pi \sqrt{2v_F^3 \kappa}}\sqrt{\sqrt{\Omega^2+E_q^2}+E_q},
\label{Pi0}
\end{align}
and
\begin{align}
E_q=v_F\left(q_x -\kappa \frac{q_y^2}{4}\right).
\end{align}
The full dynamical spin susceptibility is
\begin{align}
\chi({\bf q} + {\bf Q},\Omega)=\frac{\chi_0}{{\bf q}^2+\Pi({\bf q}+ {\bf Q},\Omega)}.
\label{th_1}
\end{align}

We now use $\chi({\bf q},\Omega)$ from  (\ref{th_1}) and calculate one-loop self-energy of an electron. We will see that relevant ${\bf q}^2$
 and $\Pi({\bf q}+ {\bf Q},\Omega)$ are of the same order, i.e., the  spin-plolarization operator evaluated at one-loop order is not a small perturbation of the bare static $\chi_0 ({\bf q}+ {\bf Q})$. Higher-loop terms in $\Pi({\bf q}+ {\bf Q},\Omega)$ are, however,  small in $\lambda$ and can be treated perturbatively.

\begin{figure}[htbp]
\includegraphics[width=.7\columnwidth]{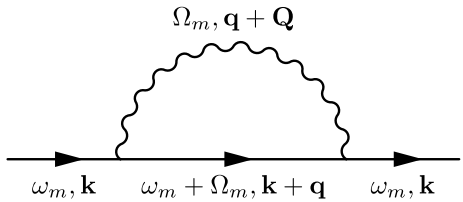}
\caption{One-loop electron self-energy.}
\label{selfen}
\end{figure}

  The one loop self-energy is presented diagrammatically in FIG. \ref{selfen}.

Specifically,
close to the hot spot, labeled by 1, the analytical expression is,
\begin{align}
&\Sigma({\bf k}, \omega_m)=\nonumber\\
&-\frac{3\bar{g}}{8\pi^3}\int
\frac{d{\bf q}~d\Omega_m}{i(\omega_m+\Omega_m)-\epsilon_{{\bf k}+{\bf q}+{\bf Q}}}\frac{1}{{\bf q}^2+ \Pi({\bf q} + {\bf Q},\Omega_m)},
\label{th_1_1}
\end{align}
where, as before, the momenta ${\bf k}$ and ${\bf k+ q}$ are taken close to hot spots 1 at $(-k_F, 0)$, and 4 at $(k_F,0)$,
 and the dispersion $\epsilon_{{\bf k}+{\bf Q}}$ is
 \beq
\epsilon_{{\bf k}+{\bf Q}} = -v_F \left(-k_x + \kappa \frac{k_y^2}{2}\right) = \epsilon_{-{\bf k}}.
  \eeq
The coefficient 3 comes from summation over the 
$x, y$, and $z$ components of the spin susceptibility.

 For definiteness, we
 consider the self-energy right at the QCP, when $\xi^{-1} =0$.
  We assume and then verify that FL behavior is preserved at the QCP, i.e. at small $\omega_m$, $\Sigma (0, \omega_m) = i \lambda \omega_m$
   and $\Sigma (k_x, 0) \propto v_F k_x$,  $\Sigma (k_y, 0) \propto  k^2_y$, We will need both frequency and momentum-dependent components of the self-energy for the calculation of $T_c$.

   To compute $\Sigma$, it is convenient to subtract from $\Sigma({\bf k}, \omega_m)$ its expression at ${\bf k} =0$ and $\omega_m=0$ as the latter vanishes by symmetry for our approximate form of the dispersion $\epsilon_k$. This subtraction can actually be done even if one does not approximate $\epsilon_k$, as, in the most general case, $\Sigma(0,0)$ accounts only for the renormalization of the chemical potential by
    fermions with energies  above $\Lambda$, which we have to neglect anyway to avoid double counting as such
     renormalization is already included  into $\epsilon_k$.  After the subtraction, the self-energy becomes
\begin{align}
\Sigma({\bf k},\omega_m)=
& \frac{3\bar{g}}{8\pi^3}\int d{\bf q}~d\Omega_m\frac{1}{{\bf q}^2+\Pi({\bf q} + {\bf Q},\Omega_m)}\nonumber\\
\times& \frac{ i\omega_m -\left[\epsilon_{-(q+k)} - \epsilon_{-q}\right]}{(i\Omega_m-\epsilon_{-q}) \left[i(\omega_m + \Omega_m) - \epsilon_{-(q+k)}\right]},
\label{th_2}
\end{align}
It is tempting to set $\omega_m$ and $k$ to zero in the denominator of (\ref{th_2}) but this would lead to an incorrect result as the integrand in (\ref{th_2})
 contains two close poles, and the contribution from the region between the poles is generally of order one~\cite{acs}. To obtain  the correct self-energy one has to explicitly integrate over internal frequency and momenta without setting $\omega_m$ and $k$ to zero.
The order of integration doesn't matter because
 the integral is convergent in the
ultra-violet limit.
 We choose to integrate over $q_x$ first
 as this will allow us to make a comparison with hole-doped cuprates.

The integral in (\ref{th_2}) above can be simplified if we introduce the dimensionless small  parameter
\begin{align}
\lambda \equiv \left(\frac{{\bar g \kappa}}{2\pi v_F}\right)^{1/3} = \left(\frac{{\bar g}}{4\pi E_F}\right)^{1/3}
\end{align}
 and rescale
\bea
&&\Sigma\rightarrow \frac{{\bar g} {\tilde \Sigma}}{\pi\lambda}, ~~  q\rightarrow\frac{\bar{g}\tilde q}{v_F\pi \lambda},~~ k\rightarrow\frac{\bar{g}\tilde k}{v_F\pi \lambda} \nonumber \\
&& \Omega_m\rightarrow\frac{\bar{g}{\tilde\Omega}_m}{\pi \lambda},~~\omega_m\rightarrow\frac{\bar{g}{\tilde\omega}_m}{\pi \lambda},
\label{an_1}
\eea
\begin{widetext}
In the new variables,
\begin{align}
&~~~~~~~~~~~~\epsilon_{q,k} = \frac{\bar g}{\pi \lambda} {\tilde \epsilon}_{{\tilde q}.{\tilde k}},~~~~{\bf q}^2 =  \left(\frac{\bar g}{\pi v_F \lambda}\right)^2 {\tilde {\bf q}}^2,  \nonumber \\
\Pi \left({\bf q} + {\bf Q}, \Omega_m\right) =\left(\frac{\bar g}{\pi v_F \lambda}\right)^2&\left[ \tilde\Pi^{(0)}\left({\tilde q_x}, {\tilde q}_y, {\tilde \Omega}_m\right)   +\tilde\Pi^{(0)}\left(-{\tilde q_x}, {\tilde q}_y, {\tilde \Omega}_m\right)+\tilde\Pi^{(0)}\left({\tilde q_y}, {\tilde q}_x, {\tilde \Omega}_m\right)+\tilde\Pi^{(0)}\left(-{\tilde q_y}, {\tilde q}_x, {\tilde \Omega}_m\right)\right]
\label{th_12}
\end{align}
\end{widetext}
where
\begin{align}
 {\tilde \epsilon}_{{\tilde q}} =&- {\tilde q}_x -  \lambda^2 {\tilde q}^2_y, \nonumber \\
 {\tilde \Pi}^{(0)} \left({\tilde q_x}, {\tilde q}_y, {\tilde \Omega}_m\right) =&\frac{1}{2}
\sqrt{\sqrt{{\tilde \Omega}^2_m + {\tilde E}^2_{\tilde q}} + {\tilde E}_{\tilde q}}, \nonumber \\
{\tilde E}_{\tilde q} =&  {\tilde q}_x -  \frac{\lambda^2 {\tilde q}^2_y}2.
\label{th_14}
\end{align}
Substituting these expressions into (\ref{th_2}) we find that the rescaled self-energy is the function of a single parameter $\lambda$:
\begin{align}
&{\tilde \Sigma} ({\bf {\tilde k}}, \omega_m)=\frac{3\lambda}{8\pi^2}\int \frac{d{\tilde q}_x~d{\tilde q}_y~d{\tilde\Omega}_m}{{\bf \tilde q}^2+\sum_{a = \pm 1}{\tilde \Pi}_{a}^{(0)}({\bf \tilde q}, {\tilde \Omega}_m)} \times\nonumber \\
& \frac{i{\tilde \omega}_m-{\tilde k}_x +\lambda^2 ({\tilde k}^2_y + 2 {\tilde k}_y {\tilde q}_y)}{(i{\tilde\Omega}_m- {\tilde q}_x)\left[i({\tilde\Omega}_m + {\tilde \omega}_m)- {\tilde q}_x - {\tilde k}_x +\lambda^2 ({\tilde k}^2_y + 2 {\tilde k}_y {\tilde q}_y)\right]},
\label{th_3}
\end{align}
where ${\tilde \Pi}_{a}^{(0)}({\bf {\tilde q}}, {\tilde \Omega}_m)=\frac12\sqrt{\left({\tilde \Omega}^2_m+{\tilde q}_x^2\right)^{1/2} +a {\tilde q}_x}+\frac12\sqrt{\left({\tilde \Omega}^2_m+{\tilde q}_y^2\right)^{1/2} + a {\tilde q}_y}$. In (\ref{th_3}) we shifted ${\tilde q}_x$ by $ \lambda^2 {\tilde q}^2_y$ and dropped all irrelevant $\lambda^2$ terms. We, however, keep ${\tilde k}^2_y$ term as it accounts for the renormalization of the
 FS curvature.

 The integrand in (\ref{th_3}), viewed as a function of ${\tilde q}_x$, contains two closely located poles coming from fermionic Green's functions, and the poles and branch cuts coming from spin susceptibility.  The two contributions can be separated as the first one comes from
  small ${\tilde q}_x$  of order ${\tilde \omega}_m$ (and  ${\tilde \Omega}_m \sim {\tilde \omega}_m$), while the one from poles and branch cuts in
$\chi ({\bf q}, \Omega_m)$ comes from  ${\tilde q}_x$ of order one. We label first contribution as  ${\tilde \Sigma}_1$ and the second as ${\tilde \Sigma}_2$.

To separate the two contributions it is convenient to divide the magnetic susceptibility into two parts as
\begin{widetext}
\begin{align}
&\frac{1}{{\bf \tilde q}^2+\sum_{a = \pm 1}{\tilde \Pi}_{a}^{(0)}({\bf \tilde q}, {\tilde \Omega}_m)} = \frac{1}{{\tilde q}_y^2+\sum_{a = \pm 1}{\tilde \Pi}_{a}^{(0)}({\tilde q}_x=0, {\tilde q}_y, {\tilde \Omega}_m)}\nonumber\\
&+ \left[\frac{1}{{\bf \tilde q}^2+\sum_{a = \pm 1}{\tilde \Pi}_{a}^{(0)}({\bf \tilde q}, {\tilde \Omega}_m)} -\frac{1}{{\tilde q}^2_y+\sum_{a = \pm 1}{\tilde \Pi}_{a}^{(0)}({q_x=0,q_y, {\tilde \Omega}_m})}\right]
\label{chu_1}
\end{align}
The pole contribution ${\tilde \Sigma}_1$ comes from the first term in the r.h.s. of (\ref{chu_1}), the branch cut contribution ${\tilde \Sigma}_2$
comes from the second term.

  The expression for ${\tilde \Sigma}_1$ is
  \begin{align}
\Sigma_1 ({\bf k}, \omega_m)&=\frac{3{\lambda}}{8\pi^2}
\int d {\tilde q}_y \int d\tilde\Omega_m \int d{\tilde q}_x   \frac{i{\tilde \omega}_m - {\tilde k}_x +\lambda^2 ({\tilde k}^2_y + 2 {\tilde k}_y {\tilde q}_y)}
{(i{\tilde\Omega}_m- {\tilde q}_x)\left[i({\tilde\Omega}_m + {\tilde \omega}_m)- {\tilde q}_x - {\tilde k}_x +\lambda^2 ({\tilde k}^2_y + 2 {\tilde k}_y {\tilde q}_y)\right]}  \nonumber \\
& \times \frac{1}{{\tilde q}^2_y+\sum_{a = \pm 1}{\tilde \Pi}_{a}^{(0)}({{\tilde q}_x=0,{\tilde q}_y, {\tilde \Omega}_m})}.
\label{th_3_1}
\end{align}
\end{widetext}
The evaluation of ${\tilde \Sigma}_1$ is straightforward -- the integral over $d {\tilde q}_x$ comes from
  the region where the two poles are in different half-planes of complex ${\tilde q}_x$. This happens when ${\tilde \Omega}_m$ is sandwiched
   between $-{\tilde \omega}_m$ and zero. Then both ${\tilde q}_x$ and ${\tilde \Omega}_m$ are small, and one can safely set
   ${\tilde \Omega}_m =0$ in the polarization operator.  The remaining integration is straightforward,
   and restoring to original variables we obtain
  \be
\Sigma_1 =  i c_1\lambda\omega_m
\label{th_3_a}
\ee
where
\begin{align}
c_1=\frac{3}{2\pi}\int_{0}^{\infty}d\tilde q_y\frac{1}{\tilde q_y^2+\sqrt{\tilde q_y/2}} =\frac{2^{4/3}}{3^{1/2}} = 1.45
\label{th_4}
\end{align}
We call this part a ``{\it non-perturbative}" contribution, because it comes from internal ${\tilde \Omega}_m$ and ${\tilde q}_x$
 comparable to external ${\tilde \omega}_m$, i.e., one cannot obtain this term by expanding in ${\tilde \omega}_m$.
 Observe that the non-perturbative contribution only depends on ${\tilde \omega}_m$, but not on ${\bf k}$.  If this would be the full self-energy, then the effective mass $m^*$ and quasiparticle residue $Z$ would be simply
  related as $Z m^*/m =1$ as in ET.

We see from (\ref{th_3}) that the non-perturbative contribution to the self-energy remains finite at the QCP and, moreover, is small as long as $\lambda$ is small.  The non-divergence of
 $d\Sigma_1/d\omega_m$ at the QCP is the consequence of including umklapp processes into $\Pi$, along with direct processes. Out of four terms in $\Pi^{(0)}_{a}$ the ones with $ a {\tilde q}_x$ under the square-root are direct processes and the ones with $a {\tilde q}_y$ are umklapp processes. When  ${\tilde q}_x = {\tilde \Omega}_m =0$, the direct component of $\tilde\Pi_{a}^{(0)}$ vanishes, and if we would keep only this term, we would obtain that the integral over $q_y$ in (\ref{th_4}) diverges as $\int d{\tilde q}_y/{\tilde q}^2_y$ and has to be cut by external $\omega_m$. In this situation,
  the non-perturbative self-energy would scale as $\omega^{3/4}_m$ (Refs.\cite{aim_el,krotkov}).  Umklapp
 processes add another contribution to $\tilde\Pi_{a}^{(0)}$, which behaves as $\sqrt{2|{\tilde q}_y|}$, and the presence of such term makes the integral in (\ref{th_4}) infra-red convergent~\cite{subir_el}.

The second  contribution to self-energy, ${\tilde \Sigma}_2$, comes from poles and branch cuts in the bosonic propagator.
We dub the contribution as ``{\it perturbative}" because typical internal ${\tilde q}_x$ and  ${\tilde \Omega}_m$ for this term are much larger than external ${\tilde \omega}_m$ and ${\tilde {\bf k}}$, hence one can safely expand in external momentum and frequency.
We have
%
\begin{widetext}
\begin{align}
{\tilde \Sigma}_2({\bf {\tilde k}},{\tilde \omega}_m)=&\frac{3{\lambda}}{8\pi^2}
\int d{\tilde q}_y~\int d{\tilde\Omega}_m \int d{\tilde q}_x  \frac{i{\tilde \omega}_m - {\tilde k}_x +\lambda^2 ({\tilde k}^2_y + 2 {\tilde k}_y {\tilde q}_y)}  {(i{\tilde\Omega}_m- {\tilde q}_x)\left[i({\tilde\Omega}_m + {\tilde \omega}_m)- {\tilde q}_x - {\tilde k}_x +\lambda^2 ({\tilde k}^2_y + 2 {\tilde k}_y {\tilde q}_y)\right]}  \nonumber \\
 \times&\left[\frac{1}{{\bf \tilde q}^2+\sum_{a = \pm 1}{\tilde \Pi}_{a}^{(0)}({\bf \tilde q}, {\tilde \Omega}_m)} -\frac{1}{{\tilde q}^2_y+\sum_{a = \pm 1}{\tilde \Pi}_{a}^{(0)}({q_x=0,q_y, {\tilde \Omega}_m})}\right] ,
\label{th_33_2}
\end{align}
Expanding in $\tilde\omega_m$, $\tilde k_x$ and $\tilde k_y$ we obtain
\begin{align}
{\tilde \Sigma}_2({\bf {\tilde k}},{\tilde \omega}_m)=&\frac{3{\lambda}}{8\pi^2}
\int d{\tilde q}_y~\int d{\tilde\Omega}_m \int d{\tilde q}_x \left[\frac{i{\tilde \omega}_m - {\tilde k}_x +\lambda^2 {\tilde k}^2_y}
{(i{\tilde\Omega}_m+ {\tilde q}_x)^2} -\frac{4\lambda^4 {\tilde q}^2_y{\tilde k}^2_y}{(i{\tilde\Omega}_m+ {\tilde q}_x)^3}\right] \nonumber \\
\times&\left[\frac{1}{{\bf \tilde q}^2+\sum_{a = \pm 1}{\tilde \Pi}_{a}^{(0)}({\bf \tilde q}, {\tilde \Omega}_m)} -\frac{1}{{\tilde q}^2_y+\sum_{a = \pm 1}{\tilde \Pi}_{a}^{(0)}({q_x=0,q_y, {\tilde \Omega}_m})}\right] ,
\label{th_3_2}
\end{align}
\end{widetext}
One can easily make sure that the 3D integral converges in the infra-red and ultra-violet limits, hence all three integrals can be taken  from minus to plus infinity.
The term with $1/(i{\tilde\Omega}_m+ {\tilde q}_x)^3$ vanishes after integration over ${\tilde q}_x$ and ${\tilde \Omega}_m$ because it is odd in these variables, but the  term with $1/(i{\tilde\Omega}_m+ {\tilde q}_x)^2$ yields a finite contribution.  Restoring to original variables, we obtain
 \be
 \Sigma_2=c_2\lambda\left[i\omega_m - v_F \left(k_x - \kappa \frac{k^2_y}{2}\right) \right],
 \label{fr_1}
 \ee
 where
 \begin{align}
 &c_2 = \frac{3}{8\pi^2} \int d{\tilde q}_y~d{\tilde\Omega}_m \int\frac{d{\tilde q}_x} {(i{\tilde\Omega}_m+ {\tilde q}_x)^2}\times\nonumber\\
 &\left[\frac{1}{{\bf \tilde q}^2+\sum_{a = \pm 1}{\tilde \Pi}_{a}^{(0)}({\bf \tilde q}, {\tilde \Omega}_m)} -\frac{1}{\tilde q^2_y+\sum_{a = \pm 1}{\tilde \Pi}_{a}^{(0)}({ \tilde q}_y, {\tilde \Omega}_m)}\right]
 \end{align}
 The numerical evaluation of  the integral yields
  \be
 c_2 = -0.75
 \ee
 Combining $\Sigma_1$ and $\Sigma_2$, we finally obtain that
  in
   hot region 1,
 \be
 \Sigma ({\bf k}, \omega_m) = 0.7 \lambda i\omega_m +0.75 \lambda v_F \left(k_x - \kappa \frac{k^2_y}{2}\right)
 \ee
 Obviously, $\partial \Sigma/\partial \omega$, $(1/v_F) \partial \Sigma/\partial k_x$ and  $(1/v_F) \partial \Sigma/\partial k^2_y$ are of the same order, and all three components of the self-energy have to be kept.

 Substituting $\Sigma ({\bf k}, \omega_m)$ into the Green's function $G^{-1} ({\bf k}, \omega_m) = i\omega_m -\epsilon_k + \Sigma (k_x, \omega_m)$ we find that at hot region 1,
 \be
 G ({\bf k}, \omega_m)= \frac{Z}{i\omega_m + v^*_F (k_x + \kappa^*\frac{k^2_y}{2})}
 \ee
  where, to first order in
  $\lambda$,
 \be
 Z = 1- 0.7 \lambda, ~~ v^*_F = v_F \frac{m}{m^*} =  1+0.05 \lambda,~~\kappa^* = \kappa(1-1.45 \lambda)
 \label{fr_2}
 \ee
We see that the dominant effect of the self-energy is the renormalization of the quasiparticle residue $Z$ and the renormalization of the curvature $\kappa$.  The  renormalization of the Fermi velocity is much smaller.

 The imaginary part of the self-energy has been calculated in Ref. [\onlinecite{subir_el}].
 At low frequencies it  scales as $\omega^{3/2}\log \omega$. The frequency dependence is stronger than in a ``conventional" FL, but still,
  ${\rm Im} \Sigma (k, \omega) \ll \omega/Z$ at small enough frequencies, hence quasiparticles near the FS remain well defined.

\section {The pairing problem}

The straightforward way to analyze whether a fermionic system  becomes superconducting below some $T_c$ is to introduce an infinitesimally
 small pairing vertex  $\Phi^{(0)}_{\alpha\beta}(k)\psi_\alpha(k)\psi_\beta(-k)$,  where $k$ stands for a three-component vector $({\bf k},\omega_m)$, renormalize it by the pairing interaction, and verify whether the
  pairing susceptibility diverges at some $T$.  The divergence of susceptibility at some $T = T_c$ implies that, below this temperature, the system is unstable against a spontaneous generation of a non-zero $\Phi_{\alpha\beta}(k)$, even  if we set $\Phi^{(0)}_{\alpha\beta}(k) =0$.
  For spin-singlet superconductivity, the spin dependence of the pairing vertex is $\Phi_{\alpha\beta} (k) = i \sigma^{y}_{\alpha \beta}\Phi (k)$.

  To obtain $T_c$ with logarithmic accuracy at weak coupling (small $\lambda$), one can restrict with only ladder diagrams for  $\Phi (k)$.
   Each additional ladder insertion contains $a \lambda \log \Lambda/T$, where $a = O(1)$.
    Ladder series are geometrical, and  summing them up one obtains $T_c \sim \Lambda e^{-a/\lambda}$. More efforts are required, however, to get the prefactor. Which diagrams have to be included depends on what theory is applicable. In Eliashberg-type  theories, all non-ladder diagram have additional smallness (in $\omega_D/E_F$ for electron-phonon interaction) and can be neglected. In this situation,
     one still can restrict with ladder diagrams, but has to solve for the full dynamical  $\Phi (k)$  beyond logarithmical accuracy, and also include fermionic self-energy to order $\lambda$.

      As an example, consider momentary Eliashberg  theory for the pairing by a single Einstein phonon.  The attractive electron-phonon interaction
  depends on transferred frequency $\Omega$ as $\lambda /[1 + (\Omega/\omega_D)^2]$. At  small $\lambda$, the normal state
   self-energy is $\Sigma = i\lambda \omega_m$,
    and the frequency dependence of the pairing vertex can be approximated as $\Phi (k) = \Phi ({\bf k}, \omega) = \Phi_0/[1+ (\omega/\omega_D)^2]$ (see Appendix A). Summing up ladder series, one then obtains \cite{phonons_1,phonons_3}, up to corrections $O(\lambda)$
    \beq
T_c = 1.13 e^{-1/2} \omega_D e^{-\frac{1+\lambda}{\lambda}} = 0.25 \omega_D e^{-\frac{1}{\lambda}}
\eeq
This result, rather than frequently cited BCS expression $T_c = 1.13 \omega_D e^{-\frac{1}{\lambda}}$, is the correct $T_c$ for weak electron-phonon interaction.

In our case  Eliashberg parameter is of order one, and we have to include on equal footing (a) ladder diagrams, which have to be taken beyond logarithmic accuracy by including the frequency dependence of the interaction, (b) the renormalization of quasiparticle $Z$, $v_F$ and $\kappa$, (c) vertex correction to the spin polarization bubble $\Pi ({\bf q+Q}, \Omega_m)$, and
(d) Kohn-Luttinger-type exchange renormalization of the pairing interaction, which in our case include vertex corrections to spin-mediated pairing interaction and exchange diagram with two crossed spin-fluctuation propagators.
   To order $O(\lambda)$, which we will need to get the prefactor in $T_c$, these four contributions add up and  can be evaluated independently.
 On the other hand we  do not need to substract from the gap equation the contribution with $k_F =0$, as it was done in Ref. [\onlinecite{gorkov-melik}],
  and add the substracted part to the renormalization of the coupling ${\bar g}$  into the scattering amplitude.  Such contributions come from energies
   above the upper cutoff of our low-energy theory, $\Lambda$, and are already incorporated into the spin-fermion coupling ${\bar g}$, which, by construction, incorporates all renormalizations from fermions with energies larger than $\Lambda$.  In momentum space, the scale $\Lambda$ roughly corresponds to $|k-k_F| \sim k_F$, but can be smaller.

 \subsection {Ladder diagrams}
\begin{figure}[htbp]
\includegraphics[width=.7\columnwidth]{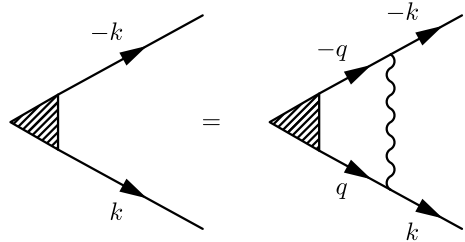}
\caption{Ladder diagrams for the pairing vertex. The wavy line denotes the interaction mediated by spin fluctuations.}
\label{ladder}
\end{figure}

We begin with ladder diagrams.  We consider spin-singlet  pairing between fermions with $k$ and $-k$,  located in opposite hot regions along the same diagonal (see FIG. \ref{ladder}).
 Because the pairing vertex is a spin singlet, $\Phi_{\alpha\beta}(k)=i\sigma_{\alpha\beta}^{y}\Phi(k)$. We denote
 by $\Phi_0 (k)$ and $\Phi_Q (k)$ the pairing vertices with momenta near ${\bf k}$ and ${\bf k} + {\bf Q}$, respectively, and treat $k$ as a small deviation from the corresponding hot spot.  Each ladder diagram
  renormalizes the pairing vertex by $\int_q G(q) G(-q) \chi (k-q + Q)$, where $\sum_{q}=T\sum_{\omega'_m}\int d^2q/(2\pi)^2$ and $Q = ({\bf Q}, 0)$ in 3D notations.  The
  ladder diagrams are readily summed up and at $T=T_c$ give rise to the integral  equation
  for $\Phi (k)$ in the form
 \begin{align}
\Phi_0(k)=-3g^2\sum_{q}\Phi_{\bf Q}(q)G(q)G(-q)\chi(k-q+Q).
\label{518_1}
\end{align}
where $\chi$ is given by Eq. (\ref{th_1}).
The overall factor $3$ comes from the convolutions of Pauli matrices at each vertex $\sigma_{\alpha'\beta'}^{y}\sigma_{\alpha\alpha'}^{i}\sigma_{\beta\beta'}^{i}=-3\sigma_{\alpha\beta}^{y}$, and the  overall minus sign
 reflects
the repulsive nature of the interaction. Superconducting instability is then possible only when $\Phi_0 (k)=-\Phi_{\bf Q}(k)$.
  The spin singlet nature of pairing requires $\Phi (k)$ to be even function of the actual 2D momentum, which in our notations implies that
   $\Phi_0({\bf k}, {\omega_m})=\Phi_{\bf Q}(-{\bf k}, {\omega_m})$.  Combined with $\Phi_0({\bf k}, {\omega_m})=-\Phi_{\bf Q}({\bf k}, {\omega_m})$, this requires $\Phi_0 ({-\bf k}, {\omega_m})=-\Phi_0 ({\bf k}, {\omega_m})$ and the same for $\Phi_Q$.
     Obviously then, the pairing amplitude passes through zero along the diagonals, i.e., the pairing symmetry is
      $d$-wave. For simplicity, below we replace  $\Phi_0 (k)$ by $\Phi (k)$ and $\Phi_{\bf Q} (k)$ by $-\Phi (k)$ and treat
       $\Phi (k)$ as an odd function of momentum (which, we remind, is counted from the corresponding hit spot along the diagonal).
       With this substitution, there will be no minus sign in the r.h.s. of the gap equation.

To simplify the calculation, we introduce the same set of rescaled dimensionless momenta and frequencies as before [Eq. (\ref{an_1})],
 and also rescale the temperature:
\begin{align}
q\rightarrow\frac{\bar{g}\tilde q}{v_F\pi \lambda},~~~{\omega_m}\rightarrow\frac{\bar{g}\tilde\omega_m}{\pi \lambda},~~~
\tilde T=\frac {\pi T\lambda}{\bar{g}}
\end{align}
 In the new variables we have,  instead of (\ref{518_1}),
\begin{align}
&\Phi({\bf \tilde k},\tilde\omega_m)=\nonumber\\
&\frac{3\lambda}{4\pi}  \tilde T\sum_{\tilde\omega_m'}\int\frac{d{\bf \tilde q}}{\tilde\omega_m'^2+\tilde q_x^2}\chi^{(0)}({\bf \tilde k} - {\bf \tilde q},\tilde\omega_m -\tilde\omega_m')\Phi({\bf \tilde q},\tilde\omega_m')
\label{32}
\end{align}
where
\begin{align}
&\chi^{(0)} ({\bf \tilde k} - {\bf \tilde q} ,\tilde\omega_m -\tilde\omega_m')= \nonumber \\
&\frac{1}{({\bf \tilde k-\tilde q})^2+\sum_{a=\pm1}\tilde\Pi^{(0)}_{a}({\bf \tilde k-\tilde q}, {\tilde \omega_m}-{\tilde\omega_m}')}.
\label{chi0}
\end{align}
As we did before,  we have shifted ${\tilde q}_x$ by $ \lambda^2 {\tilde q}^2_y$  in (\ref {32}) and dropped
$\lambda^2$ terms in the spin susceptibility $\chi^{(0)}$.
To simplify the presentation, below we will drop the tilde from the intermediate momentum and frequencies.

 We focus our attention 
 on the 
  pairing between fermions in regions 1 and 4. 
 The 
  corresponding 
  pairing vertex is an odd function of $k_y$ (the momentum component along the FS at a hot spot). For small $k_y$, we approximate the pairing vertex by $\Phi({\bf k}, {\omega'_m})=k_y\Phi(k_x,{\omega'_m})$.
  The only other place in (\ref{32}) where the dependence on $k_y$ is present is the spin susceptibility. However, it depends only on
   the relative momentum transfer ${\bf k} - {\bf q}$  and is even function of the latter. In this situation, the integration over $q_y$ gives the result proportional to $k_y$, consistent with our approximation that $\Phi({\bf k}, {\omega_m})=k_y\Phi(k_x,{\omega_m})$. In explicit form, we have 
\begin{align}
&\Phi(k_x,{\omega_m})=\nonumber\\
&\frac{3\lambda}{4\pi} \tilde T\sum_{\omega_m^\prime}\int\frac{dq_x~\chi^{(0)}(k_x-q_x,\omega_m-\omega_m^\prime)}{\omega_m^{\prime 2}+q_x^2}\Phi(q_x,\omega_m^\prime)
\label{pairing}
\end{align}
where
\begin{align}
&\chi^{(0)}(k_x -q_x,{\omega_m} -\omega_m^\prime)=\nonumber\\
&\int\frac{dk_y}{k_y^2+(k_x-q_x)^2+\sum_{a=\pm1}\tilde\Pi^{(0)}_{a}(k_x-q_x, k_y, {\omega_m}-\omega_m^\prime)}.
\label{chi}
\end{align}

The function $\chi^{(0)}(k_x -q_x, \omega_m -\omega_m^\prime)$ plays the role of the pairing kernel.
Like for electron-phonon systems, it tends to a finite value when $k_x-q_x$ and $\omega_m - \omega'_m$ vanish:
\begin{align}
\chi^{(0)}\equiv\int_{-\infty}^{\infty} \frac{dk_y}{k_y^2+\sqrt{|k_y|/2}}=\frac{2^{10/3}\pi}{3^{3/2}}=6.09.
\end{align}
To obtain the exponential term in $T_c$, we can just pull this constant out of $\tilde T\sum_{\omega_m^\prime}\int dq_x$ and evaluate the rest to
 logarithmical accuracy.  We obtain
 \beq
 T_c \propto e^{-\frac{4\pi}{3\lambda \chi^{(0)}}} = e^{-\frac{0.6874}{\lambda}}.
 \eeq

 To find the contribution from the ladder diagram to the prefactor, we use the same strategy as for electron-phonon case (see Appendix A) and write
\begin{align}
\chi^{(0)}(k_x-q_x&, \omega_m- \omega_m^\prime)=\nonumber\\
&\chi^{(0)}(k_x,{\omega_m})+\delta\chi^{(0)}(k_x,q_x;\omega_m,\omega_m^\prime),
\end{align}
where $\delta\chi^{(0)}(k_x,0;\omega_m,0)=0$.
Substituting into  Eq. (\ref{pairing}), we obtain
\begin{align}
&\Phi(k_x,{\omega_m})=\frac{3\lambda}{4\pi}\tilde T\sum_{\omega_m^\prime}\int dq_x \frac{1}{\omega_m^{\prime 2}+q_x^2} \times \nonumber \\
& {\left[\chi^{(0)}(k_x,{\omega_m})+\delta\chi^{(0)}(k_x,q_x;\omega_m,\omega_m^\prime)\right]}\Phi(\omega_m^\prime,q_x)
\label{3}
\end{align}
 Of the two terms in the last line, the first one  contains  ${\lambda}\log \tilde T\sim O(1)$, while the second one (with $\delta\chi^{(0)}(k_x,q_x;\omega_m,\omega_m^\prime)$) is convergent in the infra-red limit and is of order of  $\lambda$. The structure on the r.h.s. is reproduced on the l.h.s. if we take the pairing vertex in the form
\begin{align}
&\Phi( k_x,{\omega_m})=\Phi^0\left[ \chi^{(0)}(k_x,{\omega_m})+ \lambda \delta\Phi(k_x,{\omega_m})\right]
\label {pert}
\end{align}
Substituting this form back into the Eq. (\ref{pairing}) for $\Phi$, we obtain
\begin{widetext}
\begin{align}
&\chi^{(0)}(k_x,{\omega_m})+ \lambda \delta\Phi(k_x,{\omega_m}) \nonumber\\
&= \frac{3\lambda}{4\pi} \tilde T\sum_{\omega_m^\prime}\int dq_x~\frac{\chi^{(0)}(k_x,\omega_m)+ \delta\chi^{(0)}(k_x,q_x;\omega_m,\omega_m^\prime)}{q^2_x + \omega_m^{\prime 2}}~ \left[\chi^{(0)}(q_x,\omega_m^\prime)+ \lambda \delta\Phi(q_x,{\omega^\prime_m})\right]  \nonumber \\
& =
\frac{3\lambda^2}{4\pi} \tilde T\sum_{\omega_m^\prime}\int dq_x \frac{\chi^{(0)}(k_x,\omega_m)}{q^2_x + \omega_m^{\prime 2}} \delta\Phi(q_x,{\omega^\prime_m})+\frac{3\lambda}{4\pi} \tilde T\sum_{\omega_m^\prime}\int dq_x \frac{\chi^{(0)}(q_x,\omega_m^\prime)\chi^{(0)}(k_x-q_x,\omega_m-\omega_m^\prime)}{q^2_x + \omega_m^{\prime 2}}+ O(\lambda^2)
\label{T0}
\end{align}
To order $\lambda$ we then have
\begin{align}
& \delta\Phi(k_x,\omega_m) =  \chi^{(0)}(k_x,\omega_m)\left[P+R(k_x,\omega_m)\right],
\label{519_1}
\end{align}
where
\begin{align}
P &= \frac{3\lambda}{4\pi}\tilde T\sum_{\omega_m'}\int dq_x \frac{\delta\Phi(q_x,{\omega_m'})}{q^2_x + \omega_m^{'2}} ,\label{PP}\\
R(k_x,\omega_m)&=\frac{3}{4\pi} \tilde T\sum_{\omega_m^\prime}\int dq_x \frac{\chi^{(0)}(q_x,\omega_m^\prime)\chi^{(0)}(k_x-q_x,\omega_m-\omega_m^\prime)}{\chi^{(0)}(k_x,\omega_m)\left(q^2_x + \omega_m^{\prime 2}\right)}-\frac{1}{\lambda}\label{QQ}
\end{align}
Substituting Eq. (\ref{519_1}) into the r.h.s. of Eq. (\ref{PP}) we obtain
\begin{align}
& \left[1 - \frac{3\lambda}{4\pi} \tilde T\sum_{\omega^\prime_{m}}\int dq_x \frac{\chi^{(0)}(q_x,\omega_m^\prime)}{q^2_x + \omega_m^{\prime 2}} \right] P =  \frac{3\lambda}{4\pi}\tilde T\sum_{\omega_m'}\int dq_x \frac{\chi^{(0)}(q
_x,{\omega_m'})}{q^2_x + \omega_m^{'2}}R(q_x,{\omega_m'})
\label{519_2}
\end{align}
\end{widetext}
We now use the fact that $P = O(1)$, while the expression in the bracket of l.h.s. is of order $O(\lambda)$. The l.h.s. of (\ref{519_2}) is then of $O(\lambda)$.
 The  r.h.s. is $R(0,0)(1+O(\lambda))$. Obviously then $R(0,0)=O(\lambda)$, i.e.,
\begin{align}
&\frac{3\lambda\chi^{(0)}}{4\pi} \tilde T\sum_{\omega_m^\prime}\int dq_x \frac{\left[\chi^{(0)}(q_x,\omega_m^\prime)/\chi^{(0)}\right]^2}{\left(q^2_x + \omega_m^{\prime 2}\right)}\nonumber\\
&=1+O(\lambda^2)
\label{519_3}
\end{align}
Evaluating the integral over $q_x$ and the sum over Matsubara frequencies, we obtain
\beq
\frac{4\pi}{3\chi^{(0)}} = \lambda \log\frac{0.00874}{\tilde T_c}+O(\lambda^2)
\eeq
or, in original notations, $T_c = T_{c1}$ from ladder diagrams is
\begin{align}
 T_{c1}=0.00249~\frac{\bar g}{\lambda}~ e^{-\frac{0.6874}{\lambda}}
 \label{ttcritical}
\end{align}

\subsection {The effect of fermionic self-energy}

The self-energy corrections to ladder diagrams can be easily incorporated because in the FL regime their only role is to renormalize
 the quasiparticle $Z$, the Fermi velocity $v_F$, and the FS curvature $\kappa$.  All three renormalizations can be absorbed into the renormalization of $\lambda$ \begin{align}
\lambda=\left(\frac{\bar g\kappa}{2\pi v_F}\right)^{1/3}\rightarrow\left(\frac{\bar gZ^2\kappa^*}{2\pi v^*_F}\right)^{1/3}=\lambda(1-0.95\lambda).
\label{519_4}
\end{align}
Substituting this renormalization into  Eq. (\ref{ttcritical}) we obtain
\begin{align}
 T_{c2}=0.00130~\frac{\bar g}{\lambda}~ e^{-\frac{0.6874}{\lambda}}
 \label{ttcritical_a}
\end{align}

If Eliashberg theory was applicable to our problem, this would be the full result for $T_c$.  However, as we already discussed,
 in our case Eliashberg parameter is of order one, and other renormalizations also play a role.
 Specifically, there are two extra contributions: from vertex corrections to the polarization operator and from Kohn-Luttinger renormalization of the irreducible pairing interaction. The two contributions add up and we consider them separately.

\subsection{Correction to $T_c$ due to the renormalization of the polarization operator}

The exponential factor in the expressions for $T_{c1}$ and $T_{c2}$ is proportional to the integral
\begin{align}
\chi^{(0)} &= \int dq_y \chi (q_x=0, q_y, \omega_m =0) \nonumber\\
&= \int dq_y \frac{1}{q^2_y + \sum_{a=\pm1}\tilde\Pi^{(0)}_{a}(0, q_y, 0)}
\end{align}
 In evaluating this integral, we used the free-fermion form of the polarization operator, in which case $\sum_{a=\pm1}\tilde\Pi^{(0)}_{a}(0, q_y, 0) = \sqrt{|q_y|/2}$ and $\chi^{(0)} =6.09$.  However, to get the prefactor in $T_c$, we need to know the exponential factor with accuracy $O(\lambda)$.
 Self-energy contributions to $\tilde\Pi^{(0)}_{a}(0, q_y, 0)$ are incorporated into the renormalization of $\lambda$ in (\ref{519_4}) and are accounted for in Eq. (\ref{ttcritical_a}). However, the
   vertex correction to $\tilde\Pi^{(0)}_{a}(0, q_y, 0)$ also contributes the term of order $\lambda$, and this term has to be included.

 \begin{figure}[htbp]
\includegraphics[width=.35\columnwidth]{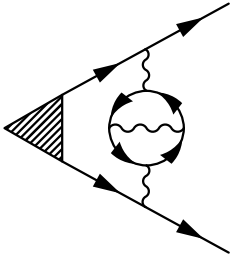}
\caption{Ladder diagrams with the effective interaction which includes vertex correction to the polarization bubble.}
\label{ladder_ver}
\end{figure}

 The effective pairing interaction with vertex correction to the polarization bubble included is shown in FIG. \ref{ladder_ver}.
  For a generic ${\bf q}$ and $\Omega_m$,  the computation of the vertex renormalization is rather messy~\cite{dan}. For our purposes, however,
 we will only need vertex correction for a static polarization bubble at ${\bf q} = (0,q_y)$, and only for {\it umklapp} process
 (i.e., in our case,  the contribution to  $\Pi$ from virtual fermions in hot region 3 and 2,

We present the computation of the vertex correction to $\Pi$ in Appendix B and here state the result -- this renormalization changes $\chi^{(0)}$ by
an $O(\lambda)$ term:
\begin{align}
\chi^{(0)}\rightarrow\chi^{(0)}(1-0.042\lambda).
\label{chi*}
\end{align}
Including this renormalization into the expression for $T_c$, Eq. (\ref{ttcritical}), we find
\begin{align}
 T_{c3}=0.00126~\frac{\bar g}{\lambda}~ e^{-\frac{0.6874}{\lambda}}
 \label{ttcritical_1}
\end{align}

\subsection{Kohn-Luttinger type corrections to effective interaction}

\begin{figure*}[htbp]
\includegraphics[width=1.8\columnwidth]{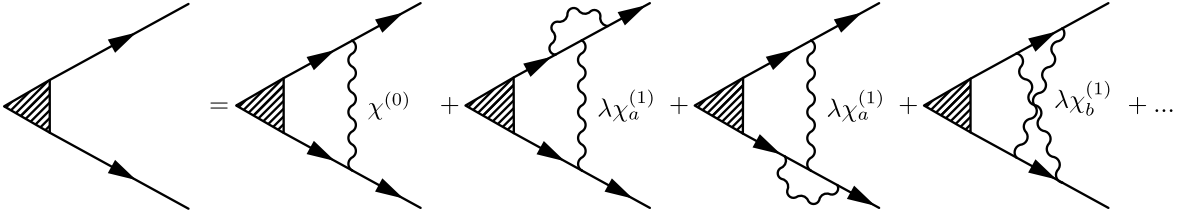}
\caption{Kohn-Luttinger diagrams for the irreducible pairing interaction. The diagram with internal particle-hole bubble is already included into $\chi^{(0)} (k)$ and has to be dropped to avoid double counting.  Each of the remaining three Kohn-Luttinger diagrams gives contribution of order $\lambda$.}
\label{fig1}
\end{figure*}

Finally, we consider the effect on $T_c$ from  Kohn-Luttinger type second-order corrections to the effective pairing interaction.
 We show the corresponding diagrams in FIG. \ref{fig1}.  Compared to the original Kohn-Luttinger work \cite{kl}, we have dropped one diagram since in our case it is already included into $\chi^{(0)} (k)$.

Similar to what we did before  with the dynamical part of the interaction in ladder series, we analyze the equation for the pairing vertex using the full $\chi$ which we split into the original and the Kohn-Luttinger terms.
We have

\begin{align}
&\Phi({\bf  k},\omega_m)=\frac{3\lambda}{4\pi}  \tilde T\sum_{\omega_m'}\int\frac{d{\bf q}}{\omega_m'^2+ q_x^2}\chi({\bf  k},\omega_m ; {\bf  q},\omega_m')\Phi({\bf q},\omega_m')
\label{519_5}
\end{align}
where
\begin{align}
\chi ({\bf  k},\omega_m ; {\bf  q},\omega_m')=& \chi^{(0)} (k_x-q_x,k_y-q_y,\omega_m-\omega_m')\nonumber\\
+&{\lambda}\chi^{(1)} (k_x,k_y,\omega_m;q_x,q_y,\omega_m');
\end{align}
 $\chi^{(0)} (k-q)$, defined in Eq. (\ref{chi0}), accounts for the first graph in FIG. \ref{fig1},
 and ${\lambda}\chi^{(1)} (k,q)$ accounts for the other three terms.

The first term in $\chi (k,q)$  depends on the momentum transfer $k_y-q_y$. For this reason we could shift, in the r.h.s. of (\ref{519_4}), the integration over $q_y$ by the integral over $q_y- k_y$ and obtain the pairing vertex as a linear function of $k_y$.
This simple scaling with $k_y$ does not extend to Kohn-Luttinger terms because $\chi^{(1)} (k,q)$ depends separately on $k$ and on $q$.
 Accordingly, we write
\begin{align}
\Phi(k_x,k_y,{\omega_m})=\Phi^0[k_y\chi^{(0)}(k_x,{\omega_m})+\lambda \delta\Phi(k_x,k_y,{\omega_m})].
\label{519_6}
\end{align}
Because we are interested in $O(\lambda)$ terms, we can safely neglect the difference between continuous and discrete Matsubara frequencies and treat
$\omega_m$ as continuous variable.

Plugging (\ref{519_6})  back to the pairing equation and formally setting  $k_x={\omega_m}=0$, we obtain
\begin{widetext}
\begin{align}
k_y\chi^{(0)}+ \lambda\delta\Phi(0,k_y,0)&=\frac{3\lambda}{4\pi}  \tilde T\sum\int\frac{dq_x}{({\omega^\prime_m})^{2}+q_x^2}k_y\left(\chi^{(0)}(q_x,\omega_m^\prime)\right)^2\nonumber\\
&+\frac{3\lambda^2}{4\pi}  \tilde T\sum\int\frac{dq_x}{({\omega^\prime_m})^{2}+q_x^2}\int dq_y\chi^{(1)}(0,k_y,0;q_x,q_y,\omega_m^\prime)q_y\chi^{(0)}(q_x,\omega_m^\prime)\nonumber\\
&+\frac{3\lambda^2}{4\pi}  \tilde T\sum\int\frac{dq_x}{({\omega^\prime_m})^{2}+q_x^2}\int dq_y\chi^{(0)}(q_x,k_y-q_y,\omega_m^\prime)\delta\Phi(q_x,q_y,\omega_m^\prime)\nonumber\\
\end{align}
Evaluating the integrals,  we obtain
\begin{align}
k_y\chi^{(0)}+\lambda\delta\Phi(0,k_y,0)&=k_y \frac{3\lambda}{4\pi} \chi^{(0)2} \log{\frac{0.00874}{{\tilde T}}}
+\frac{3\lambda^2}{4\pi}  \chi^{(0)}\log\frac{a}{\tilde T}\int dq_y \chi^{(1)}(0,k_y,0;0,q_y,0)q_y\nonumber\\
&+\frac{3\lambda^2}{4\pi}  \log\frac{{b}}{\tilde T}\int dq_y \chi^{(0)}(0,k_y-q_y,0)\delta\Phi(0,q_y,0)+O(\lambda^2).
\label{519_7}
\end{align}
\end{widetext}
 where $a$ and $b$ are constants of order one.


Simplifying the notations and rearranging, we re-express (\ref{519_7}) as
\begin{align}
\frac{4\pi}{3\lambda \chi^{(0)}}= \log{\frac{0.00874}{\tilde T}}
+\frac{A(k_y) + C(k_y)}{k_y(\chi^{(0)})^2}+O(\lambda^2).
\label{519_8}
\end{align}
where we  defined
\begin{align}
A(k_y) &= \int dq_y \left[\frac{\chi^{(0)}(k_y-q_y)}{\chi^{(0)}} \delta\Phi(0,q_y,0)\right]-\delta\Phi(0,k_y,0)\nonumber\\
C(k_y)&=\int dq_y q_y \chi^{(1)}(0,k_y,0;0,q_y,0)
\label{xi}
\end{align}
and $\chi^{(0)}(k_y-q_y)=1/[(k_y-q_y)^2+\sqrt{|k_y-q_y|/2}]$.
One can easily make sure that
$A(0)=C(0)=0$, hence $[A(k_y)+\xi^{(1)}(k_y)]/k_y$ is not singular when $k_y$ vanishes.  However,
Eq. (\ref{519_7}) sets a more stringent requirement:  $A(k_y)+C(k_y)$ must be equal to $B k_y (\chi^{(0)})^2$, where $B$ is a constant.
  Then $T_c = T_{c3} e^B$, where $T_{c3}$ is given by Eq. (\ref{ttcritical_1}).

  The  $A(k_y)$ in Eq. (\ref{xi})  contains $\delta\Phi (0,k_y,0)$ and the integral over $q_y$ of
$\delta\Phi (0,y_y,0)$, weighted  with a kernel.  The condition
\begin{align}
A(k_y)+C(k_y) = B k_y (\chi^{(0)})^2
\label{ABC}
\end{align}
then sets the integral equation on $\delta \Phi (0,k_y,0)$.
 We solve this equation in Appendix C and obtain $B$ in the form
  $B=C(-\Lambda)/(\chi^{(0)})^2$, where $\Lambda$ is the dimensionless momentum cutoff along the FS.

 The value of $B$ depends on the interplay between $\lambda$ and $1/\Lambda$.
 For a generic FS, the momentum cutoff $\Lambda$ is of order $k_F$, in which case $\lambda \Lambda \sim (W/{\bar g})^{1/3} \gg 1$, where $W$ is fermionic bandwidth. In this situation,  the pairing interaction dies off at momenta $k_y$, which are parametrically smaller than the cutoff, and
  and $C(-\Lambda)$ is small in $1/(\lambda \Lambda)$. Then $B$ is also small in $1/(\lambda \Lambda)$, hence $e^B \approx 1$,  i.e., the correction to $T_c$ from Kohn-Luttinger diagrams can be neglected. In this situation,  the fully renormalized $T_c$
  coincides with  $T_{c3}$  and is given by
 \begin{align}
 T_{c}=0.00126~\frac{\bar g}{\lambda}~ e^{-\frac{0.6874}{\lambda}}
 \label{ttcritical_2}
\end{align}
Equation (\ref{ttcritical_2}) is the main result of this paper.

 If, by some reasons, $\Lambda$ is numerically much smaller than $k_F$, such that  $\lambda \Lambda$ is actually a small number, the renormalization of $T_c$ due to Kohn-Luttinger diagrams becomes relevant. We present the  calculation of $B$ for this case in Appendix C.  We find that,  at small $\lambda \Lambda$,
 $B$ is logarithmically singular:  $B = (10/3) \log{1/(\lambda \Lambda)} + ...$ where the ellipsis stand for  terms $O(1)$. As a result,
  $T_c$ is enhanced by the factor $(1/\lambda \Lambda)^{10/3}$  compared to Eq. (\ref{ttcritical_2}). The outcome is that
  Eq.  (\ref{ttcritical_2}) provides the lower boundary for $T_c$ -- the actual $T_c$ gets enhanced by Kohn-Luttinger contributions. How strong the enhancement is depends on the actual band structure, which set the value of $\Lambda$.

 \subsection{Comparison with the experiments on electron-doped cuprates}

We now compare our theoretical $T_c$, Eq. (\ref{ttcritical_2}), with the data for near-optimally doped ${\rm Nd}_{2-x}{\rm Ce}_x{\rm Cu}_3{\rm O}_4$~ and ${\rm Pr}_{2-x}{\rm Ce}_x{\rm Cu}_3{\rm O}_4$,  in which doping creates extra electrons.
The parameters of the quasiparticle dispersion, $v_F$ and $\kappa$, can be extracted from the  ARPES measurements on
${\rm Nd}_{2-x}{\rm Ce}_x{\rm Cu}_3{\rm O}_4$~\cite {exp_el}.  We found $v_F=0.87 \rm eV$ and $\kappa=0.31$ (in units where the lattice constant $a=1$). Similar parameters have been obtained in ~\cite{mar_fit}.
The only other input parameter for the theory is the strength of spin-fermion coupling ${\bar g}$.
For hole-doped cuprates, the fits to ARPES and NMR data in the normal state yielded ${\bar g} \leq 2 \rm eV$ (Ref. [\onlinecite{acs}]).
This ${\bar g}$ is consistent with the value of the charge-transfer gap in the effective Hubbard model in the Mott-Heisenberg regime at half-filling~\cite{opt_1/2_hole} as calculations in the ordered state of the spin-fermion model~\cite{morr} place the gap to be exactly ${\bar g}$ -- quantum corrections cancel out.  This consistency is not an anticipated result as
 ${\bar g}$ extracted from the optics is the coupling at high-energies, comparable to $E_F$, while the one used in the comparison with ARPES and NMR is the coupling at low-energies (below our $\Lambda$), where, strictly speaking, spin-fermion model is only valid.  The (rough) agreement between the two likely implies that renormalizations between $E_F$ and $\Lambda$ do reduce ${\bar g}$, but only by a small fraction.
 For electron-doped cuprates,  the detailed fits of $\Sigma (k, \omega)$ in the spin-fermion model to the self-energy,  extracted from ARPES data, have not been done yet, but  optical measurements~\cite{optics_el} show that
    the charge-transfer gap is about $1.7\rm eV$.  Assuming that the situation in electron-doped cuprates is  the same as in hole-doped cuprates, i.e., that
     spin-fermion couplings,  extracted from ARPES and optics, are not far each other, we  just take ${\bar g}$ to be equal to this $1.7\rm eV$.

 Using the numbers for ${\bar g}$, $v_F$, and $\kappa$, we obtain $\lambda \sim 0.46$, which implies that weak coupling analysis should be applicable.
     Substituting $\lambda =0.46$ and ${\bar g} \approx 20000 \rm K$, we find
      $T_c \sim  0.0006 {\bar g}  \sim 12 \rm K$. This is reasonably close to the experimental
     $T_c = 20-24 \rm K$  in optimally doped
     ${\rm Nd}_{2-x}{\rm Ce}_x{\rm Cu}_3{\rm O}_4$~ \cite{takagi,TC2} particularly given that our theoretical $T_c$, Eq. (\ref{ttcritical_2}) is the lower boundary
       for the actual $T_c$ because (i) as we found above, $T_c$ goes up once we include corrections due to a finite upper cutoff of the theory, and (ii)
       in real situation, hot spots at optimal doping are still located at some distance from each other, in which case the value of $T_c$ should move a bit towards the one when hot spots are well separated, and the latter is much higher:
        when hot spots are near $(0,\pi)$ and symmetry-related points, $T_c \sim 0.04 {\bar g}$   ~\cite{wang,norman_1}.

The transition temperature in the similar range of $10-20\rm K$  has been found in FLEX calculations~\cite{FLEX}, and the agreement between our and FLEX results in an encouraging sign.  The authors of ~\cite{mar_fit,bansil} considered the model with a static  interaction $V$, extracted $V$ by fitting the value of the magnetization in the antiferromagnetically ordered state, and used BCS formula for $T_c$. Amazingly, their $T_c$ is quite similar to the one we obtained.
 The two-particle-self-consistent approach, applied to the Hubbard model with nearest-neighbor hopping only and values of the interaction $U$ typical of electron-doped systems, yields a much higher optimal $T_c \sim 200\rm K$ [Refs. \onlinecite{tremblay,X}]. However, given the sensitivity of $T_c$ to the specifics of the FS \cite{XX}, the value of $T_c$  in this approach has to be reanalyzed using the model for the hopping consistent with the measured FS.

\section{Summary}

In this work, we  re-visited the issue of normal state renormalizations and superconducting $T_c$ in electron-doped cuprates near optimal doping.
We used spin-fermion model to model electronic interactions and assumed that the doping at which magnetic order with ${\bf Q}=(\pi, \pi)$ sets in is close to the one at which the Fermi surface touches the magnetic Brillouin zone boundary along the zone diagonals (this case is often labeled as ${\bf Q}=2{\bf k}_F$).

 Quantum-critical fluctuations and the pairing instability in the ${\bf Q}=2{\bf k}_F$ case have been studied before~\cite{aim_el,krotkov}. However, recent work~\cite{subir_el} has shown that earlier analysis did not include umklapp processes and, as a result, severely overestimated the strength of quantum-critical fluctuations.  Once umklapp process are properly accounted for, the real part of the self-energy at a QCP scales as $\omega$ and the imaginary part  behaves  as $\omega^{3/2} \log{\omega}$, i.e., fermionic coherence is preserved at the lowest energies.

 The goal of this work was to re-visit the calculation of $T_c$. We found that the argument~\cite{krotkov} that the  $d_{x^2-y^2}$ superconductivity survives  when hot spots
 merge along Brillouin zone diagonals, holds.  However, the value of $T_c$ has to be re-considered.
   The calculation of $T_c$ requires one to know
  the renormalization of the quasiparticle propagator  in the normal state, and in the first part of the paper we computed the real part of the fermionic self-energy (which was not considered in Ref. \cite{subir_el}). We found that  Eliashberg approximation is not valid because the Eliashberg parameter is of order one, and the only way to proceed with calculations is to perform a direct perturbative loop expansion.
   We found that Re $\Sigma (k, \omega)$ is a regular function of momentum and frequency,
 and the renormalizations of the quasiparticle residue $Z$, Fermi velocity $v_F$, and the FS curvature $\kappa$ hold on powers of the  single dimensionless parameter $\lambda$. We  treated  $\lambda$ as small parameter and obtained $Z$, $v_F$, and $\kappa$  to order $O(\lambda)$.

We then used normal state results as an input and computed superconducting $T_c$ by solving the (2+1)-dimensional  gap equation in momentum and frequency.
  To logarithmical accuracy, the solution of the linearized gap equation  is similar to that in BCS theory, and $T_c \propto e^{-a/\lambda}$, where in our case $a = 0.6874$. We, however, computed $T_c$ with the  prefactor, which required us to go one step beyond BCS approximation and include the frequency dependence of the interaction, the renormalizations of $Z$, $v_F$, and $\kappa$, vertex corrections to particle-hole polarization bubble, and Kohn-Luttinger (non-ladder) corrections to the irreducible
 pairing interaction. Using the parameters extracted from the data on optimally electron-doped cuprates,  we found $T_c \geq 10\rm K$,
   which is in a reasonably good agreement with the experimental values. The agreement is particularly striking because our result is the
   lower boundary of the actual $T_c$.

 One issue brought about by our work in comparison with earlier works on the Hubbard model~\cite{tremblay,tremblay_1,X,XX} is the origin of the difference between hole and electron-doped cuprates. The reasoning displayed in~\cite{tremblay,tremblay_1,X,XX,gabi} is  that the interaction $U$ is somewhat smaller in electron-doped cuprates than in hole-doped
  cuprates such that in electron-doped materials correlations are relevant, but Mott physics does not develop. This is certainly a valid point as, e.g., the magnetic $T_N$ is smaller in half-filled  ${\rm Pr}_{2-x}{\rm Ce}_x{\rm Cu}_3{\rm O}_4$,and ${\rm Pr}_{2-x}{\rm Ce}_x{\rm Cu}_3{\rm O}_4$ than in undoped La and Y based materials). Our results, however, point on a complementary  reason for the difference between near-optimally hole and electron-doped cuprates. Namely,
   even if interaction (our ${\bar g}$) is the same, there is still a substantial difference between the magnitude  of fermionic self-energy and of superconducting $T_c$ due to the difference in the geometry of the electronic FS.

\acknowledgments

We acknowledge useful conversations with D. Chowdhury, T. Das, R. Greene, I. Eremin, S. Maiti, S. Sachdev, and particularly A-M. Tremblay.
  We are thankful to I. Mazin for the discussion on the prefactor in the formula for $T_c$ in the weak coupling limit of the Eliashberg theory.  The work was supported by the DOE grant DE-FG02-ER46900.

\appendix

\section{$T_c$ at weak coupling in a phonon superconductor}

A portion of our calculation of $T_c$ is similar to the calculation of $T_c$ in the weak-coupling limit of Eliashberg theory for a  phonon superconductor for the case when a phonon propagator can be approximated by a single Einstein mode:
\beq
\chi_{ph} (\Omega_m) = \frac{\chi_0}{\Omega^2_m + \omega^2_D}
\label{cth_1}
\eeq
In Eliashberg theory, $T_c$ is the temperature at which the linearized equation for the pairing vertex $\Phi (\Omega_m)$ has a non-zero solution.
 The equation for $\Phi (\Omega_m)$ is well-known~\cite{phonons,phonons_1,phonons_2,phonons_3} and  in rescaled variables ${\bar T} = T/\omega_D$,
 ${\bar \omega}_m = \omega_m/\omega_D = \pi {\bar T} (2m+1)$  reads
\beq
\Phi ({\bar \Omega}_m) = \pi {\bar T} \lambda^* \sum_m \frac{\Phi ({\bar \omega}_m)}{|{\bar \omega}_m|} \chi(\bar\omega_m,\bar \Omega_m)
\label{cth_2}
\eeq
 where $\chi(\bar\omega_m,\bar \Omega_m)\equiv \frac{1}{1+\left({\bar \omega}_m - {\bar \Omega}_m\right)^2}$ and $\lambda^* = \lambda/(1 + \lambda)$  [$\lambda$ is dimensionless effective electron-phonon coupling and the factor $(1 + \lambda)$ comes from mass renormalization]
The formula for $T_c$ in Eliashberg theory at weak coupling  has been discussed several times in the past~\cite{phonons_1,phonons_2,phonons_3}.  However, until now, there is some
confusion about the interplay between the weak coupling limit of the Eliashberg theory and the BCS theory~\cite{phonons}.
Within BCS theory (extended to include $1+\lambda$ mass renormalization), the pairing vertex is approximated by a constant and the dependence on the external ${\bar \Omega}_m$ in the bosonic propagator is neglected. The equation for $T_c$ then reduces to
\beq
1 =  \lambda^* \sum_m\frac{1}{|2m+1|} \frac{1}{1 + \pi^2 {\bar T}^2 (2m+1)^2}
\label{cth_3}
\eeq
The sum in the r.h.s.  converges at the  largest $m$ and is
 expressed in terms of di-Gamma functions. At small ${\bar T}$ it reduces to $\log {1.13/{\bar T}}$.
 From (\ref{cth_4}) we then obtain, in original notations,
 \beq
 T^{BCS}_c = 1.13 \omega_D e^{- \frac{1 + \lambda}{\lambda}}
\label{cth_4}
\eeq
The point made in Refs.\cite{phonons_1,phonons_2,phonons_3} is that this expression  is {\it not} the correct $T_c$ in the small $\lambda$ limit of the Eliashberg theory.  The correct formula, obtained first in Ref. \cite{phonons_1} (see also~\cite{phonons_2,phonons_3}), is
 \beq
 T_c = 1.13 e^{-1/2} \omega_D e^{- \frac{1 + \lambda}{\lambda}} = 0.69 \omega_D e^{- \frac{1 + \lambda}{\lambda}}
\label{cth_5}
\eeq
The reason for the discrepancy between Eqs. (\ref{cth_4}) and (\ref{cth_5}) is that in Eliashberg theory
 the numerical prefactor in $T_c$ comes from fermions with
  energies of order $\omega_D$ (${\tilde \omega}_m = O(1)$), and for such fermions the dependence of the pairing vertex $\Phi ({\bar \omega}_m)$
   on ${\bar \omega}_m$  cannot be neglected.

   The computational procedure presented in Ref.~\cite{phonons_1} and in subsequent work~\cite{phonons_3} uses iteration method and is somewhat involved.
   Below we present an alternative computation procedure to obtain  Eq. (\ref{cth_5}). We use the same procedure in the calculations
    of $T_c$ for our case of electron-doped cuprates.
    We re-express $\chi(\bar\omega_m,\bar \Omega_m)\equiv \frac{1}{1+\left({\bar \omega}_m - {\bar \Omega}_m\right)^2} $
    in Eq. (\ref{cth_2}) as
    \begin{align}
    \chi(\bar\omega_m,\bar \Omega_m)&\equiv\chi(\bar \Omega_m)+\delta\chi(\bar\omega_m,\bar\Omega_m)\nonumber\\
    \chi(\bar \Omega_m)&= \frac{1}{1 + {\bar \Omega}^2_m}\nonumber\\
    \delta\chi(\bar\omega_m,\bar\Omega_m)&= \frac{\bar \omega_m}{1 + {\bar \Omega}^2_m} \frac{2  {\bar \Omega}_m - {\bar \omega}_m}{1 + \left({\bar \omega}_m - {\bar \Omega}_m\right)^2}.
    \label{cth_6}
    \end{align}
    Plugging this expression back to Eq. (\ref{cth_2}) we find that the first term in Eq. (\ref{cth_6}) gives $\lambda^*\log\bar T\sim 1$, while the second term is free of logarithm and is of order $\lambda^*$. We then search for the solution of Eq. (\ref{cth_2}) in the form
    \beq
    \Phi ({\bar \Omega}_m) = {\Phi_0} \left[\chi(\bar\Omega_m)+\lambda^*\delta\Phi(\bar\Omega_m)\right]+O(\lambda^{*2}),
   \label{cth_7}
    \eeq
where $\delta \Phi ({\bar \Omega}_m)$ is assumed to be independent of $\lambda^*$ up to corrections of order $\lambda^*$. Substituting Eq. (\ref{cth_7})
 into Eq. (\ref{cth_2}) and neglecting non-logarithmical terms of order $(\lambda^*)^2$, we obtain
 \beq
 \delta \Phi ({\bar \Omega}_m) = \chi({\bar \Omega}_m)\left[ P + R({\bar \Omega}_m)\right],
   \label{cth_8}
    \eeq
    where
    \begin{align}
    P &=  \lambda^*\pi T_c \sum \frac{\delta \Phi ({\bar \omega}_m)}{|{\bar \omega}_m|}
 \label{cth_90}\\
    R({\bar \Omega}_m) &= \frac{\pi T_c}{\chi(\bar\Omega_m)} \sum  \frac{1}{|{\bar \omega}_m|} \chi(\bar\omega_m)\chi(\bar\omega_m,\bar\Omega_m)-\frac{1}{\lambda^*}
     \label{cth_9}
    \end{align}
Observe that Eq. (\ref{cth_8}) is not an integral equation on $\delta \Phi ({\bar \Omega}_m)$ because the integral term $P$ does not
  depend on ${\bar \Omega}_m$.

Constructing $P$  in the r.h.s. of Eq. (\ref{cth_8}), we obtain
\beq
\left[1-\lambda^*\pi\bar T_c\sum\frac{\chi(\bar\omega_m)}{|\bar\omega_m|}\right]P=\lambda^*\pi\bar T_c\sum\frac{\chi(\bar\omega_m)}{|\bar\omega_m|}R(\bar\omega_m)
 \label{cth_10}
    \eeq
Because $1-\lambda^*\pi\bar T_c\sum\frac{1}{|\bar\omega_m|}\chi(\bar\omega_m) = O(\lambda^*)$ and $P$ is at most of order $O(1)$, the l.h.s. of Eq. (\ref{cth_10}) is $O(\lambda^*)$. On the other hand, the r.h.s of Eq. (\ref{cth_10}) becomes
\begin{align}
\lambda^*\pi\bar T_c\sum\frac{1}{|\bar\omega_m|}\chi(\bar\omega_m)R(\bar\omega_m) =R(0) \left(1 +O(\lambda^*)\right).
\end{align}
Matching both sides, we find that
\begin{align}
R(0)=O(\lambda^*),
\end{align}
hence
\beq
\lambda^*\pi T_c \sum  \frac{1}{|{\bar \omega}_m|} \frac{1}{\left(1 + {\bar \omega}^2_m\right)^2} = 1+O(\lambda^{*2})
 \label{cth_11}
    \eeq
Evaluating the sum (it is expressed in terms of di-Gamma functions) and taking the limit ${\bar T} \ll 1$, we reproduce
Eq. (\ref{cth_5}).

Another way to obtain this result is to formally set  $\bar\Omega_m=0$ in Eq. (\ref{cth_8}), which gives $\delta\Phi(0)=P+R(0)$. At the same time,   from  Eq. (\ref{cth_90}) we have $P=\delta\Phi(0)+O(\lambda^*)$. Matching the two expressions, we reproduce $R(0)=O(\lambda^*)$.

\section{Evaluation of the contribution to $T_c$ from vertex corrections to the polarization bubble}

 \begin{figure}[htbp]
\includegraphics[width=196.8pt]{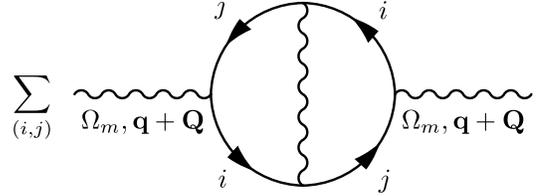}
\caption{Vertex correction to polarization bubble. Labels $i$ and $j$ denotes fermions at different hot spots. The full expression is the sum of
  direct and umklapp processes, $(1,4), (4,1)$, $(2,3), (3,2)$.}
\label{RPA_ver}
\end{figure}

In this Appendix we present the evaluation of the contribution to $T_c$  from vertex correction to the polarization bubble $\Pi^{(1)}({\bf q+Q},\Omega_m)$.

The diagram for $\Pi^{(1)}({\bf q+Q},\Omega_m)$ is shown in FIG. \ref{RPA_ver}. As we discussed, vertex correction to the polarization bubble contributes to $T_c$ by renormalizing $\chi^{(0)}$. For our purposes, one can easily make sure that we only need to include the renormalization of the
   spin susceptibility and at  transferred momentum ${\bf q}=(0,q_y)$.  For momentum along $y$ axis,  the contribution from umklapp process dominates, because in direct process $q_y$ dependence is suppressed by $\lambda$.
We denote the
contribution from umklapp process at $\left(q_x=0, q_y, \Omega_m=0\right)$ as $\Pi_{um}^{(1)}(0,q_y,0)$. We have,

\begin{widetext}
\begin{align}
\Pi_{um}^{(1)}&(0,q_y,0)=-2\bar g^2\int\frac{d{\bf k}_1~d\omega_{m1}}{(2\pi)^3}\int\frac{d{\bf k}_2~d\omega_{m2}}{(2\pi)^3}\frac{1}{\left({\bf k}_1-{\bf k}_2\right)^2+\Pi\left({\bf k}_1-{\bf k}_2,\omega_{m1}-\omega_{m2}\right)}\nonumber\\
&\times\frac{1}{i\omega_{m1}-\bar\epsilon_{k_1}}\frac{1}{i\omega_{m1}-\bar\epsilon_{-k_1+q}}\frac{1}{i\omega_{m2}-\bar\epsilon_{k_2}}\frac{1}{i\omega_{m2}-\bar\epsilon_{-k_2+q}}+\left({\bf q}\rightarrow-{\bf q}\right),
\end{align}
 where
\begin{align}
\bar\epsilon_k=v_F\left(k_y-\kappa\frac{k_x^2}{2}\right),
\end{align}
where the coefficient $(-2)$ comes from Pauli matrix algebra. Using
a set of
 rescaling variables 
 similar to the one 
 introduced in Eq. (\ref{an_1}) in the main text, we obtain
\begin{align}
\Pi_{um}^{(1)}(0,q_y,0)= \lambda \left(\frac{\bar g}{\pi v_F\lambda}\right)^2 \tilde\Pi_{um}^{(1)}(0,\tilde q_y,0),
 \end{align}
where
 \begin{align}
 \tilde\Pi_{um}^{(1)}(0,\tilde q_y,0)=&-4\pi^2\int\frac{du~dx~dy}{(2\pi)^3}\int\frac{dU~dX~dY}{(2\pi)^3}\frac{1}{X^2+Y^2+\sum_{a}\tilde\Pi_{a}^{(0)}(X,Y,U)}\nonumber\\
 \times&\frac{1}{-i\left(u-\frac{U}{2}\right)+\left(y-\frac{Y}{2}+\frac{\tilde q_y}{2}\right)-x^2}\frac{1}{i\left(u-\frac{U}{2}\right)+\left(y-\frac{Y}{2}-\frac{\tilde q_y}{2}\right)+x^2}\nonumber\\
 \times&\frac{1}{i\left(u+\frac{U}{2}\right)+\left(y+\frac{Y}{2}+\frac{\tilde q_y}{2}\right)+x^2}\frac{1}{-i\left(u+\frac{U}{2}\right)+\left(y+\frac{Y}{2}-\frac{\tilde q_y}{2}\right)-x^2}
 \label{Pi1ver}
 \end{align}


  Eq. (\ref{Pi1ver}) is a 6$D$ integral, for which numerical schemes  designed to evaluate multidimensional integrals, such as Monte-Carlo,
    do not yield satisfactory results.
   Fortunately, the integration  over $u, x$ and $y$, which are rescaled frequency and two momentum components in the
    fermionic loop, can be done analytically. The remaining integration over $U, X, Y$, can be done numerically to a good precision.

To perform the integration, we first write down
\begin{align}
\tilde\Pi_{um}^{(1)}(0,\tilde q_y,0)=-4\pi^2I(\tilde q_y),
\label{I}
\end{align}
where
\begin{align}
 I(\tilde q_y)=&\int\frac{dU~dX~dY}{(2\pi)^3}\int\frac{du~dx~dy}{(2\pi)^3}\frac{1}{X^2+Y^2+\sum_{a}\tilde\Pi_{a}^{(0)}(X,Y,U)}\nonumber\\
 \times&\underbrace{\frac{1}{-i\left(u-\frac{U}{2}\right)+\left(y-\frac{Y}{2}+\frac{\tilde q_y}{2}\right)-x^2}}_{(1)}\underbrace{\frac{1}{i\left(u-\frac{U}{2}\right)+\left(y-\frac{Y}{2}-\frac{\tilde q_y}{2}\right)+x^2}}_{(2)}\nonumber\\
 \times&\underbrace{\frac{1}{i\left(u+\frac{U}{2}\right)+\left(y+\frac{Y}{2}+\frac{\tilde q_y}{2}\right)+x^2}}_{(3)}\underbrace{\frac{1}{-i\left(u+\frac{U}{2}\right)+\left(y+\frac{Y}{2}-\frac{\tilde q_y}{2}\right)-x^2}}_{(4)}
\end{align}
We use the residue theorem for the integration over $dy$. There are four poles, from terms labeled (1) (2) (3) and (4).
 We split $I$ into two parts, $I(\tilde q_y)=I_1(\tilde q_y)+I_2(\tilde q_y)$. $I_1$ comes from the range where the poles in (1) and (4) are in the same half plane, while $I_2$ comes from the range where the poles in (2) and (4) are in the same half plane.

For $I_1$ we obtain
\begin{align}
I_1(\tilde q_y)&=2\pi i\int\frac{dU~dX~dY}{(2\pi)^3}\int_{|u|>\frac{|U|}2}du~{\rm sgn}{(u)}\int\frac{dx}{(2\pi)^3}\frac{1}{X^2+Y^2+\sum_{a}\tilde\Pi_{a}^{(0)}(X,Y,U)}\nonumber\\
&\times\frac{1}{2x^2+Y+2iu}~\frac{1}{2x^2-{\tilde q_y}+2i\left(u-\frac{U}{2}\right)}~\frac{1}{-iU+Y-{\tilde q_y}}+\left({\tilde q_y}\rightarrow-{\tilde q_y}\right)
\end{align}
rearranging the second line and combining the integrals over positive and negative $u$, we obtain
\begin{align}
I_1(\tilde q_y)&=\int\frac{dU~dX~dY}{16\pi^5}\frac{1}{X^2+Y^2+\sum_{a}\tilde\Pi_{a}^{(0)}(X,Y,U)}\int_{\frac{|U|}2}^{\infty}du\int_{-\infty}^{\infty} dx\nonumber\\
&\times{\rm Im}\left\{\left[\frac{1}{2x^2+Y+2iu}-\frac{1}{2x^2-{\tilde q_y}+2i\left(u-\frac{U}{2}\right)}\right]\frac{1}{Y^2-({\tilde q_y}+iU)^2}\right\}+\left({\tilde q_y}\rightarrow-{\tilde q_y}\right)
\end{align}
We next perform the integration over $x$ and over $u$ using the same steps as we did in the calculation of the one-loop polarization bubble in the main text. Carrying out the integrations, we obtain
\begin{align}
I_1(\tilde q_y)&=-\int\frac{dU~dX~dY}{16\pi^4\sqrt{2}}\frac{1}{X^2+Y^2+\sum_{a}\tilde\Pi_{a}^{(0)}(X,Y,U)}\nonumber\\
&\times{\rm Re}\left[\left(\sqrt{Y+i|U|}-\sqrt{-{\tilde q_y}-iU+i|U|}\right)\frac{1}{Y^2-({\tilde q_y}+iU)^2}\right]+\left({\tilde q_y}\rightarrow-{\tilde q_y}\right)
\label{an_2}
\end{align}
Folding the integration over $U$ to positive $U$, we re-write (\ref{an_2}) as
 \begin{align}
I_1(\tilde q_y)&=\int_{0}^{\infty} dU\int\frac{dX~dY}{16\pi^4\sqrt{2}}\frac{1}{X^2+Y^2+\sum_{a}\tilde\Pi_{a}^{(0)}(X,Y,U)}\nonumber\\
&\times{\rm Re}\left[\left(-2\sqrt{Y+iU}+\sqrt{{-\tilde q_y}}+\sqrt{{\tilde q_y+2iU}}\right)\frac{1}{Y^2-({\tilde q_y}+iU)^2}\right]+\left({\tilde q_y}\rightarrow-{\tilde q_y}\right)
\label{I1}
\end{align}

Similarly, for $I_2$ we obtain
\begin{align}
I_2(\tilde q_y)&=2\pi i\int_{-\infty}^{\infty}dU~{\rm sgn}{(U)}\int\frac{dX~dY}{(2\pi)^3}\int_{-\frac{|U|}2}^{\frac{|U|}{2}}du\int\frac{dx}{(2\pi)^3}\frac{1}{X^2+Y^2+\sum_{a}\tilde\Pi_{a}^{(0)}(X,Y,U)}\nonumber\\
&\times\frac{1}{2x^2-Y+2iu}~\frac{1}{2x^2-{\tilde q_y}+2i\left(u-\frac{U}{2}\right)}~\frac{1}{iU+Y+{\tilde q_y}}+\left({\tilde q_y}\rightarrow-{\tilde q_y}\right)
\end{align}
rearranging the second line, folding the integration over $U$ to positive $U$, and integrating over $x$ and $u$ we obtain
\begin{align}
I_2(\tilde q_y)&=\int_{0}^{\infty} dU\int\frac{dX~dY}{16\pi^4\sqrt{2}}\frac{1}{X^2+Y^2+\sum_{a}\tilde\Pi_{a}^{(0)}(X,Y,U)}\nonumber\\
&\times{\rm Re}\left[\left(\sqrt{Y+iU}-\sqrt{Y-iU}-\sqrt{-{\tilde q_y}}+\sqrt{{-\tilde q_y-2iU}}\right)\frac{1}{Y^2-({\tilde q_y}+iU)^2}\right]+\left({\tilde q_y}\rightarrow-{\tilde q_y}\right)
\label{I2}
\end{align}

Combining $I_1$ and $I_2$ and substituting into (\ref{I}),  we obtain
\begin{align}
\tilde\Pi_{um}^{(1)}(0,\tilde q_y,0)&=-\int_{0}^{\infty} dU\int\frac{dX~dY}{4\pi^2\sqrt{2}}\frac{1}{X^2+Y^2+\sum_{a}\tilde\Pi_{a}^{(0)}(X,Y,U)}\nonumber\\
&\times{\rm Re}\left[\left(-\sqrt{Y-iU}-\sqrt{Y+iU}+\sqrt{{-\tilde q_y-2iU}}+\sqrt{{\tilde q_y+2iU}}\right)\frac{1}{Y^2-({\tilde q_y}+iU)^2}\right]+\left({\tilde q_y}\rightarrow-{\tilde q_y}\right)
\label{Pi111}
\end{align}
\end{widetext}

\begin{figure}[htbp]
\includegraphics[width=246.0pt]{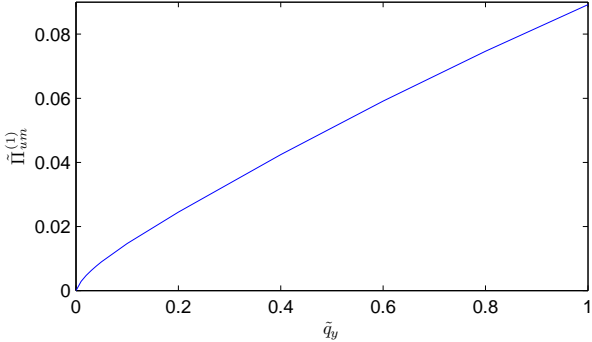}
\caption{The dependence of $\tilde\Pi_{um}^{(1)}(0,\tilde q_y,0) - \tilde\Pi_{um}^{(1)}(0,0,0)$  on $\tilde q_y$.}
\label{Pium}
\end{figure}

   The integration over $U, X, Y$ has been done numerically.
    Just as we did in the calculation of the one-loop polarization operator, we subtract from $\tilde\Pi_{um}^{(1)}(0,\tilde q_y,0)$ its value at zero momentum $\tilde\Pi_{um}^{(1)}(0,0,0)$ to make the integral infra-red convergent.
  The term we subtract only shift the position of the QCP and is not of interest to us.
After the subtraction, the integration in (\ref{Pi111}) can be extended to an infinite range.

We plot $\tilde\Pi_{um}^{(1)}(0,\tilde q_y,0) - \tilde\Pi_{um}^{(1)}(0,0,0)$ in FIG. \ref{Pium}. We see that numerically
 $\tilde\Pi_{um}^{(1)}(0,\tilde q_y,0) - \tilde\Pi_{um}^{(1)}(0,0,0)$ is small even when ${\tilde q}_y =1$.

The $O(\lambda)$ correction to $\chi^{(0)}$, which we need for the calculation of the right prefactor for $T_c$,
   is related to  $\tilde\Pi_{um}^{(1)}(0,\tilde q_y,0) - \tilde\Pi_{um}^{(1)}(0,0,0)$ as
\begin{align}
\chi^{(0)}\rightarrow\chi^{(0)}-\lambda\int_{-\infty}^{\infty} dz \frac{\tilde\Pi_{um}^{(1)}(0,z,0) - \tilde\Pi_{um}^{(1)}(0,0,0)}{\left(z^2+\sqrt{|z|/2}\right)^2},
\end{align}
Using the numerical results for $\tilde\Pi_{um}^{(1)}(0,\tilde q_y,0) - \tilde\Pi_{um}^{(1)}(0,0,0)$,
we found that the renormalization of $\chi^{(0)}$ by the vertex correction in the polarization operator is
\begin{align}
\chi^{(0)}\rightarrow\chi^{(0)}(1-0.042\lambda),
\end{align}
We cited this result in  Eq. (\ref{chi*}) in the main text.

\section{Contribution to $T_c$ from Kohn-Luttinger diagrams at finite momentum cutoff}
We found in the main text that Kohn-Luttinger diagrams  renormalize the transition  temperature $T_c$ to
\begin{align}
T_c = T_{c3} \exp{B},
\label{tca}
\end{align}
where $T_{c3}$, given by Eq. (\ref{ttcritical_1}), is the transition temperature without Kohn-Luttinger contributions, and $B$ is the prefactor for $k_y$ term term in the integral equation
\begin{align}
A(k_y)+C(k_y) = B k_y (\chi^{(0)})^2.
\label{ABCa}
\end{align}
where $A$ and $C$ are given by
\begin{align}
A(k_y) &= \int_{-\Lambda}^{\Lambda} dq_y\left[\frac{\chi^{(0)}(q_y)}{\chi^{(0)}} \delta\Phi(k_y+q_y)\right]-\delta\Phi(k_y)\nonumber\\
C(k_y)&=\int_{-\Lambda}^{\Lambda} dq_y q_y \chi^{(1)}(k_y,q_y),
\label{xia}
\end{align}
and we remind that $\chi^{(0)} (q_y) = 1/(q^2_y + \sqrt{|q_y|/2})$, $\chi^{(0)}$ is the number ($=6.09$), and $\chi^{(1)}(k_y,q_y)$ accounts for the contributions from Kohn-Luttinger diagrams.  We also used shorthand notations $\delta\Phi(k_y)\equiv\delta\Phi(k_x=0,k_y,\omega_m=0)$ and $\chi^{(1)}(k_y,q_y)\equiv\chi^{(1)}(k_x=0,k_y,\omega_m=0;q_x=0,q_y,\omega_m'=0)$ and we explicitly restricted the momentum integration to $|q_y|  < \Lambda$, where $\Lambda$ is the upper momentum cutoff for the low-energy theory.

To obtain $B$, we take the first and second derivatives of Eq. (\ref{ABCa}):
\begin{align}
A'(k_y)+C'(k_y)&=B(\chi^{(0)})^2,\label{1st}\\
A''(k_y)+C''(k_y)&=0.\label{2nd}
\end{align}
Integrating Eq. (\ref{2nd}) from its lower limit $-\Lambda$ to $k_y$, we find
\begin{align}
A'(k_y)+C'(k_y)&=A'(-\Lambda)+C'(-\Lambda).
\label{1st_a}
\end{align}
Hence
\begin{align}
B(\chi^{(0)})^2=A'(-\Lambda)+C'(-\Lambda).
\end{align}
Now, let's explicitly write down
\begin{align}
&A'(-\Lambda)=\int_{-\Lambda}^{\Lambda}dq_y\frac{\chi^{(0)}(q_y)}{\chi^{(0)}}\left[ \delta\Phi'(q_y-\Lambda)-\delta\Phi'(-\Lambda)\right]
\end{align}
 Typical values of $q_y$ are set by $\chi^{(0)}(q_y)$ and are order $O(1)$. Because $\Lambda\gg 1$ by construction,
 $\delta\Phi'(q_y-\Lambda)-\delta\Phi'(-\Lambda) = O(1/\Lambda)$, i.e., $A'(-\Lambda)$ is small and can be neglected.
 Then
\begin{align}
B(\chi^{(0)})^2=C'(-\Lambda)=\int_{-\Lambda}^{\Lambda} dq_y q_y \left.\frac{\partial\chi^{(1)}(k_y,q_y)}{\partial k_y}\right |_{k_y=-\Lambda}.
\label{Bchi}
\end{align}

We now need the explicit expression for $\chi^{(1)}(k_y,q_y)$.  Evaluating explicitly the three Kohn-Luttnerg contributions in FIG. \ref{fig1}, we obtain, for zero external frequency and $x$-component of momenta
 $\chi^{(1)}(k_y,q_y)=2\chi_a^{(1)}(k_y,q_y)+\chi_b^{(1)}(k_y,q_y)$, where
\begin{widetext}
\begin{align}
&\chi_a^{(1)}(k_y,q_y;\lambda)\nonumber\\
&=-\frac{1}{8\pi^2}\int\frac{du~dx~dy}{iu-x-\lambda^{2}(k_y-y)^2}\frac{1}{iu+x-\lambda^{2}(q_y-y)^2}\nonumber\\
&\times\frac{1}{x^2+y^2+\sum_{a=\pm1}\tilde\Pi_a^{(0)}(x,y,u)}\frac{1}{(k_y-q_y)^2+\sqrt{|k_y-q_y|/2}}
\label{a}
\end{align}
and
\begin{align}
&\chi_b^{(1)}(k_y,q_y;\lambda)\nonumber\\
&=-\frac{3}{8\pi^2}\int\frac{du~dx~dy}{iu-x-\lambda^{2}(k_y+y)^2}\frac{1}{iu+x-\lambda^{2}(q_y-y)^2}\nonumber\\
&\times\frac{1}{x^2+y^2+\sum_{a=\pm1}\tilde\Pi_a^{(0)}(x,y,u)}\frac{1}{x^2+(y+k_y-q_y)^2+\sum_{a=\pm1}\tilde\Pi_a^{(0)}(x,y+k_y-q_y,u)}.
\label{b}
\end{align}
\end{widetext}
The difference in prefactors comes from different structures in Pauli matrices convolution and is trivial to verify. In these two integrals the ${\lambda^2}$ terms in the denominator, together with external momenta $k_y$ and $q_y$, serves as an infra-red cutoff.

 One can easily make sure that the magnitude of $C^{\prime}(-\Lambda)$ depends on the value ${\lambda}^2\Lambda^2$. One can show that,
  if the momentum cutoff along the FS is of order $k_F$,  ${\lambda^2}\Lambda^2\sim(W/\bar{g})^{2/3}\gg 1$, where $W$ is the fermionic bandwidth. In this case we find that $B(\chi^{(0)})^2=C^{\prime}(-\Lambda)\ll 1$. Then $B \ll 1$, and  the renormalization due to Kohn-Luttinger-type diagrams are small.

If, however, the momentum cutoff $\Lambda$ is much smaller such that $\lambda\Lambda$ is actually a small number, the integral for
$C'(-\Lambda)$ in (\ref{Bchi}) contains $\log\frac{1}{\lambda\Lambda}$.  Collecting the contributions from (\ref{a}) and (\ref{b}), we obtain at
 $\lambda\Lambda \ll 1$
 \begin{align}
C'(-\Lambda)&=\frac{10}{3}\left(\chi^{(0)}\right)^2\log\frac{1}{\lambda\Lambda} + O(1)
\end{align}
  Using Eqs. (\ref{Bchi}) and (\ref{tca}), we find that $T_c$ in this case will be enhanced by a factor of $\exp B\sim\left({\lambda\Lambda}\right)^{-10/3}$.



\begin{thebibliography}{99}
\bibitem{comm_a} Strictly speaking, this decoupling is fully justified only when the theory is extended to $N \gg1$ fermionic flavors
 because the fermionic self-energy, evaluated using $\chi (q, \Omega_m)$ with Landau damping,  does contain, along with the frequency depedence, also
  the dependence on the momenta along the FS, $k_\parallel$  The latter does affect the prefactor in $\Pi (Q, \Omega) = \gamma \Omega_m$. The relative
   change in $\gamma$ compated with the case when $\Sigma ({\bf k}, \omega_m) = \Sigma (\omega_m)$ is of order  $1/N^2$ and is small at large $N$.
\end{thebibliography}
\end {document}